\pdfoutput=1
\documentclass[10pt,a4paper,ptm,twoside]{article}

\usepackage[utf8]{inputenc}
\usepackage[english]{babel}

\usepackage{amssymb}
\usepackage{amsmath}
\usepackage{amsfonts}

\usepackage{xcolor,graphicx,soul,multirow}
\usepackage[round]{natbib}
\usepackage[affil-it]{authblk}
\usepackage[outline]{contour}
\contourlength{.2pt}
\usepackage[breaklinks=true,bookmarks=true,colorlinks=true,
            linkcolor=blue,citecolor=blue,urlcolor=blue]{hyperref}
\usepackage{nomencl}
\makenomenclature

\newcommand{\figdir}{./}
\newcommand{\refdir}{./}
\usepackage{geometry}
\newlength\halflineskip
\newlength\affilskip
\usepackage{fancyhdr}
\usepackage[font=small,labelfont=sc]{caption}
\usepackage{titlesec}
\titleformat{\section}
  {\bf\large}
  {\thesection}{1em}{\vspace*{-.2cm}}
\usepackage{etoolbox}
\AtBeginEnvironment{tabular}{\small}

\usepackage{array}
\newif\ifcolmn
\colmnfalse
\newif\ifref
\reftrue
\newif\ifall
\alltrue
\newif\ifone
\newif\iftwo
\newif\ifthree
\newif\iffour
\newif\iffive
\newif\ifsix
\newif\ifseven
\newif\ifeight
\newif\ifnine
\newif\iften
\newif\ifeleven
\newif\iftwelve
\newif\ifthirteen
\newif\iffourteen
\newif\iffifteen
\newif\ifsixteen
\newif\ifseventeen
\newif\ifeighteen
\newif\ifnineteen
\newif\iftwenty
\newif\iftwentyone
\newif\iftwentytwo
\newif\iftwentythree
\newif\iftwentyfour
\newif\iftempfig
\newif\ifrevfig
\onefalse
\twofalse
\threefalse
\fourfalse
\fivefalse
\sixfalse
\sevenfalse
\eightfalse
\ninefalse
\tenfalse
\elevenfalse
\twelvefalse
\thirteenfalse
\fourteenfalse
\fifteenfalse
\sixteenfalse
\seventeenfalse
\eighteenfalse
\nineteenfalse
\twentyfalse
\twentyonefalse
\twentytwofalse
\twentythreefalse
\twentyfourfalse
\tempfigfalse
\revfigfalse
\ifall
\onetrue
\twotrue
\threetrue
\fourtrue
\fivetrue
\sixtrue
\seventrue
\eighttrue
\ninetrue
\tentrue
\eleventrue
\twelvetrue
\thirteentrue
\fourteentrue
\fifteentrue
\sixteentrue
\seventeentrue
\eighteentrue
\nineteentrue
\twentytrue
\twentyonetrue
\twentytwotrue
\twentythreetrue
\twentyfourtrue
\tempfigtrue
\revfigtrue
\fi
\twentytwotrue
\ninefalse
\twentyfourfalse
\revfigfalse
\newcommand\runa{\emt{\text{run~1}}}                           % Run nr.
\newcommand\runb{\emt{\text{run~2}}}                           % Run nr.
\newcommand\runbf{\emt{\text{run~2~(fix)}}}                    % Run nr.
\newcommand\rund{\emt{\text{run~3}}}                           % Run nr.
\newcommand\emt[1]{\ensuremath{#1}}
\newcommand\np{\emt{    N_p}}                                  % Number of particles
\newcommand\iind{\emt{   \ell}}                                % particle index
\newcommand\vot{\emt{    \mathcal{V}}}                         % Total volume
\newcommand\dy{\emt{     \Delta x_2}}                          % grid spacing 2-direction
\newcommand\Lx{\emt{     L_{x1}}}                              % Domain size 1-direction
\newcommand\Ly{\emt{     L_{x2}}}                              % Domain size 2-direction
\newcommand\Lz{\emt{     L_{x3}}}                              % Domain size 3-direction
\newcommand\nx{\emt{     N_{x1}}}                              % Grid points 1-direction
\newcommand\ny{\emt{     N_{x2}}}                              % Grid points 2-direction
\newcommand\nz{\emt{     N_{x3}}}                              % Grid points 3-direction
\newcommand\up[1]{\emt{  u_{#1}^{(p)*}}}                        % Particle velocity
\newcommand\uf[1]{\emt{  u_{#1}}}                              % Fluid velocity
\newcommand\uff[1]{\emt{    (\uf{1}^{#1},\,\uf{2}^{#1},\,\uf{3}^{#1})}} % Fluid velocity components
\newcommand\xf[1]{\emt{  x_{#1}}}                              % Cartesian coordinate
\newcommand\xb[1]{\emt{  \boldsymbol{x}^{#1}}}                 % Cartesian coordinate vector
\newcommand\xff[1]{\emt{    (\xf{1}^{#1},\,\xf{2}^{#1},\,\xf{3}^{#1})}} % Cartesian coordinates
\newcommand\var{\emt{    \psi^{(p)}}}                          % Generic Particle-Related Variable
\newcommand\varf{\emt{   \psi^{(f)}}}                                % Generic Fluid-Related Variable
\newcommand\sav[1]{\emt{ \left\langle #1\right\rangle}} % Volume average
\newcommand\zxav[1]{\emt{\left\langle #1\right\rangle}_{\Lz,\,\Lx}}  % xz-average
\newcommand\Int{\emt{    \displaystyle\int}}                   % Integral
\newcommand\Sump{\emt{   \displaystyle\sum\limits_{\iind=1}^{\np}}}% Sum over particles
\newcommand\svf{\emt{    \phi_s}}                              % Solid Volume Fraction
\newcommand\mpart{\emt{    m_p}}                               % Solid Volume Fraction
\newcommand\idf{\emt{    \varphi^{(p)}}}                       % Transfer function
\newcommand\idff{\emt{   \varphi^{(f)}}}                       % Transfer function (fluid)
\newcommand\fref[1]{\emt{  f_{ref}}}                           % Reference force
\newcommand\dudx[3]{     \emt{\dfrac{\partial #1^{#3}}{\partial \xf{#2}^{#3}}}}        % df/dx
\newcommand\stenf{\emt{  \sigma_{ij}^{(f)*}}}                   % Fluid stress tensor
\newcommand\mob{\emt{    \Psi}}           % Mobility number
\newcommand\denss{\emt{  \varrho_s}}      % Grain density
\newcommand\densf{\emt{  \varrho}}        % Fluid density
\newcommand\g{\emt{      g}}              % Gravitational acceleration
\newcommand\vs{\emt{     v_s}}            % Gravitational velocity
\newcommand\ds{\emt{     d}}              % Sphere diameter 1
\newcommand\s{\emt{      s}}              % Relative density
\newcommand\U{\emt{      U_0}}            % Free-stream velocity oscillation amplitude
\newcommand\Ue{\emt{     U_e}}            % Free-stream velocity
\newcommand\del{\emt{    \delta}}         % Stokes BL thickness
\newcommand\om{\emt{     \omega}}         % Angular frequency
\newcommand\TT{\emt{     T}}              % Oscillation period
\newcommand\Rdel{\emt{   R_\del}}         % Reynold \delta
\newcommand\Rd{\emt{     Re_\ds}}         % Reynold particle
\newcommand\Kc{\emt{     K_c}}            % Keulegan-Carpenter number
\newcommand\frifac{\emt{ f_w}}            % Friction factor
\newcommand\tautot{\emt{ \tau_{b}}}       % Total shear stress
\newcommand\taumax{\emt{ \tau_{b,max}}}       % Max total shear stress
\newcommand\taup{\emt{ \tau_{part}}}      % particle shear stress
\newcommand\tauv{\emt{ \tau_{visc}}}      % viscous shear stress
\newcommand\taut{\emt{ \tau_{turb}}}      % turbulent shear stress
\newcommand\shields{\emt{\theta}}         % Shields number
\newcommand\qpmean{\emt{q_s}}             % mean part. flowrate
\newcommand\phase{\emt{  \varphi}}        % Cycle phase
\newcommand\vort[1]{\emt{ \Omega_{#1}}}   % Vorticity components
\newcommand\avel[1]{\emt{ \omega_{#1}^{(p)*}}} % Angular velocity

\title{\bf %
Interface-resolved direct numerical simulations of sediment transport in a turbulent oscillatory boundary layer%
}%

\author[1]{{\bf MARCO MAZZUOLI~\footnote{\textit{e-mail:}~\texttt{marco.mazzuoli@unige.it}}}}
\author[1]{{\bf PAOLO BLONDEUAX}}
\author[1]{{\bf GIOVANNA VITTORI}}
\author[2]{{\bf MARKUS UHLMANN}}
\author[3]{{\bf JULIAN SIMEONOV}}
\author[3]{{\bf JOSEPH CALANTONI}}
\affil[1]{{\small Department of Civil, Chemical and Environmental Engineering (DICCA), University of Genoa, %\\[\affilskip]
Via Montallegro 1, 16145 Genova, Italy}}
\affil[2]{{\small Institute for Hydromechanics, Karlsruhe Institute of Technology,% \\[\affilskip]
76131 Karlsruhe, Germany}}
\affil[3]{{\small Marine Geosciences Division - Naval Research Laboratory - Stennis Space Center Mississipi, U.S.A.}}

\date{November 2019}

\begin{document}

\maketitle

\vspace{.5cm}

\begin{abstract}
The flow within an oscillatory boundary layer, which approximates the flow generated by propagating sea waves of small amplitude close to the bottom, is simulated numerically by integrating Navier-Stokes and continuity equations. %
The bottom is made up of spherical particles, free to move, which mimic sediment grains. %
The approach allows to fully-resolve the flow around the particles and to evaluate the forces and torques that the fluid exerts on their surface. %
Then, the dynamics of sediments is explicitly computed by means of Newton-Euler equations. %
For the smallest value of the flow Reynolds number presently simulated, the flow regime turns out to fall in the intermittently turbulent regime such that turbulence appears when the free stream velocity is close to its largest values but the flow recovers a laminar like behaviour during the 
remaining phases of the cycle. %
For the largest value of the Reynolds number turbulence is significant almost during the whole flow cycle. %
The evaluation of the sediment transport rate allows to estimate the reliability of the empirical predictors commonly used to estimate the amount of sediments transported by the sea waves. %
For large values of the Shields parameter, the sediment flow rate during the accelerating phases does not differ from that observed during the decelerating phases. %
However, for relatively small values of the Shields parameter, the amount of moving particles depends not only on the bottom shear stress but also on flow acceleration. %
Moreover, the numerical results provide information on the role that turbulent eddies have on sediment dynamics. %
\end{abstract}

\section{Introduction}
\label{introd}

In nature, flows which involve the motion of solid particles coupled to that of a fluid are quite common and different models have been developed to predict phenomena involving the motion of sediment particles and either air or water. %
The approaches employed to describe sediment and fluid motions are different depending on the spatial scale of interest which can range from a few millimetres to hundreds of kilometres, i.e. from the scale of the sediment grains to the scale of the largest morphological patterns observed on the Earth surface (e.g. tidal sand banks). %

Depending on the problem under investigation, the interstitial fluid can play a minor role in the transport of momentum and the rheology of the mixture is mainly controlled by the phenomena occurring during direct grain-grain contacts. %
On the other hand, under different conditions, as it happens in dilute suspensions, the motion of the fluid plays a primary role in the dynamics of the mixture. %
Finally, the hydrodynamic force acting on sediment grains and the force due to grain-grain contacts could be equally important, as it happens at the bottom of water bodies (seas, lakes, rivers, estuaries, ...) where flow drag can mobilise sediment grains arrested on the bed surface by gravity and frictional contacts. %

The threshold conditions for the initiation of sediment transport and the sediment transport rate are usually determined by considering the average velocity field and neglecting the turbulent fluctuations \citep[see i.a.][]{graf1984,fredsoe1992,soulsby1997,gyr2006}. %
However, the vortex structures which characterise a turbulent flow might induce local high values of the fluid velocity and mobilise the sediment particles even when the average flow is relatively weak. %
Despite a lot of experimental studies having been devoted to investigate the mechanisms responsible for the initiation of sediment transport and the complex dynamics of sediment grains, a clear and detailed picture of the interaction of coherent vortex structures and sediment particles is still missing. %

The flow generated by a monochromatic surface wave of small amplitude propagating over a flat sandy bottom provides a fair description of the actual flow that is observed in coastal environments seawards of the breaker zone. %
Close to the bottom, the surface wave induces oscillations of the pressure gradient and originates an oscillatory boundary layer (OBL). %
The OBL is characterised by \emph{(i)} the amplitude $\U^*$ of the irrotational velocity oscillations close to the bottom, \emph{(ii)} the order of magnitude $\del^*=\sqrt{2\nu^*/\om^*}$ of the thickness of viscous bottom boundary layer and \emph{(iii)} the angular frequency $\om^*=2\pi/\TT^*$ of the surface wave, where $\TT^*$ is the wave period. %
Hereinafter, a star is used to denote a dimensional quantity while the same symbol without the star denotes its dimensionless counterpart. %
Moreover, let the mechanical properties of sea water be assumed constant and represented by the density $\densf^*$ and the kinematic viscosity $\nu^*$. %
The sediments are assumed to be cohesionless, monodisperse and characterised by the density $\denss^*$ and the diameter $\ds^*$ of the grains. %

The dynamics of the OBL over the seabed is rich because features typical of the laminar, transitional and turbulent regimes might coexist during a flow cycle depending on the values of the Reynolds number $\Rdel=\U^*\del^*/\nu^*$ and the dimensionless particle diameter $\ds=\ds^*/\del^*$. %

The OBL over a smooth wall (i.e. for $\ds=0$) becomes turbulent if $\Rdel$ is larger than $550$ \citep{Costamagna2003}, while, in the presence of particles, this value progressively decreases being approximately equal to $500$ for $\ds=2.32$, $400$ for $\ds=2.80$ and $150$ for $\ds=6.95$ \citep{mazzuoli2016b,ghodke2016,ghodke2018,mazzuoli2019b}. %
Laboratory observations show that different flow regimes exist within the OBL over a smooth wall, namely, the laminar regime, the disturbed laminar regime, the intermittently turbulent regime and the fully developed turbulent regime. %
In the disturbed laminar regime, small perturbations of the Stokes flow appear but the average flow does not deviate significantly from that observed in the laminar regime. %
The intermittently turbulent regime is characterized by the appearance of turbulent bursts during the decelerating
phases of the cycle but the flow recovers a laminar like behaviour during the accelerating phases. %
Finally, in the fully developed turbulent regime, turbulence is present during the whole oscillation cycle \citep{hino1976,hino1983,jensen1989,akhavan1991,Carstensen2010}. %
Later, the  experimental observations found a theoretical interpretation by \citet{wu1992} and \citet{blondeaux1994} who showed that the appearance of turbulence is due to both nonlinear 3-D effects and a receptivity mechanism. %
These theoretical findings were later supported by the results of Direct Numerical Simulations (DNSs) of Navier-Stokes and continuity equations \citep{akhavan1991,Verzicco1996,Costamagna2003,ozdemir2014}. %
It is worth pointing out that similar results were obtained by considering a rough wall, even though the roughness of the wall makes turbulence to appear for smaller values of the Reynolds number \citep{jensen1989,carstensen2012,mazzuoli2019b}. %

Even when, for small values of both $\Rdel$ and $\ds$, the flow never becomes turbulent during the flow cycle, the prediction of the sediment transport rate is challenging because sediments are subject both to the viscous drag and to the effects of the wave-driven pressure-gradient. %
\citet{mazzuoli2019} investigated the formation of ripples in the OBL for $\Rdel=72$ and $128$ over a bed of spherical mono-sized particles of dimensionless diameter $\ds$ equal to $0.25$. %
By means of DNS, \citet{mazzuoli2019} showed that the contribution of the pressure-gradient to the particle dynamics can be significant. %
Indeed, they observed that a significant amount of sediment was mobilised also during phases characterized by small values of the bed shear stress. %
However, if the size $\ds^*$ of the sediment is sensibly smaller than $\delta^*$, this contribution can be neglected and the sediment flow rate is well correlated with the bed shear stress that is fairly approximated by that of a Stokes boundary layer. %

Laboratory experiments carried out by \citet{lobkovsky2008}, in absence of turbulent fluctuations, revealed that sediment dynamics rapidly adapts to the slow changes of the driving flow, and the sediment flow rate could be estimated, also in unsteady conditions, by a power-law function of the excess of bed shear stress with respect to the critical value for incipient sediment motion. %
\citet{mazzuoli2019} concluded that, as long as turbulence appearance is not triggered, both the viscous and pressure-gradient contributions were  mainly controlled by the parameter $\mob/\Rdel$, $\mob=\U^{*2}/(\frac{\denss^*-\densf^*}{\densf^*}\g^*\,\ds^*)$ denoting the mobility number and $\g^*$ being the module of the gravitational acceleration. %

\begin{figure}
\begin{picture}(0,185)(0,0)
  \ifone
  \put(60,0){\includegraphics[trim=0cm 0cm 0cm 0cm, clip, width=.7\textwidth]{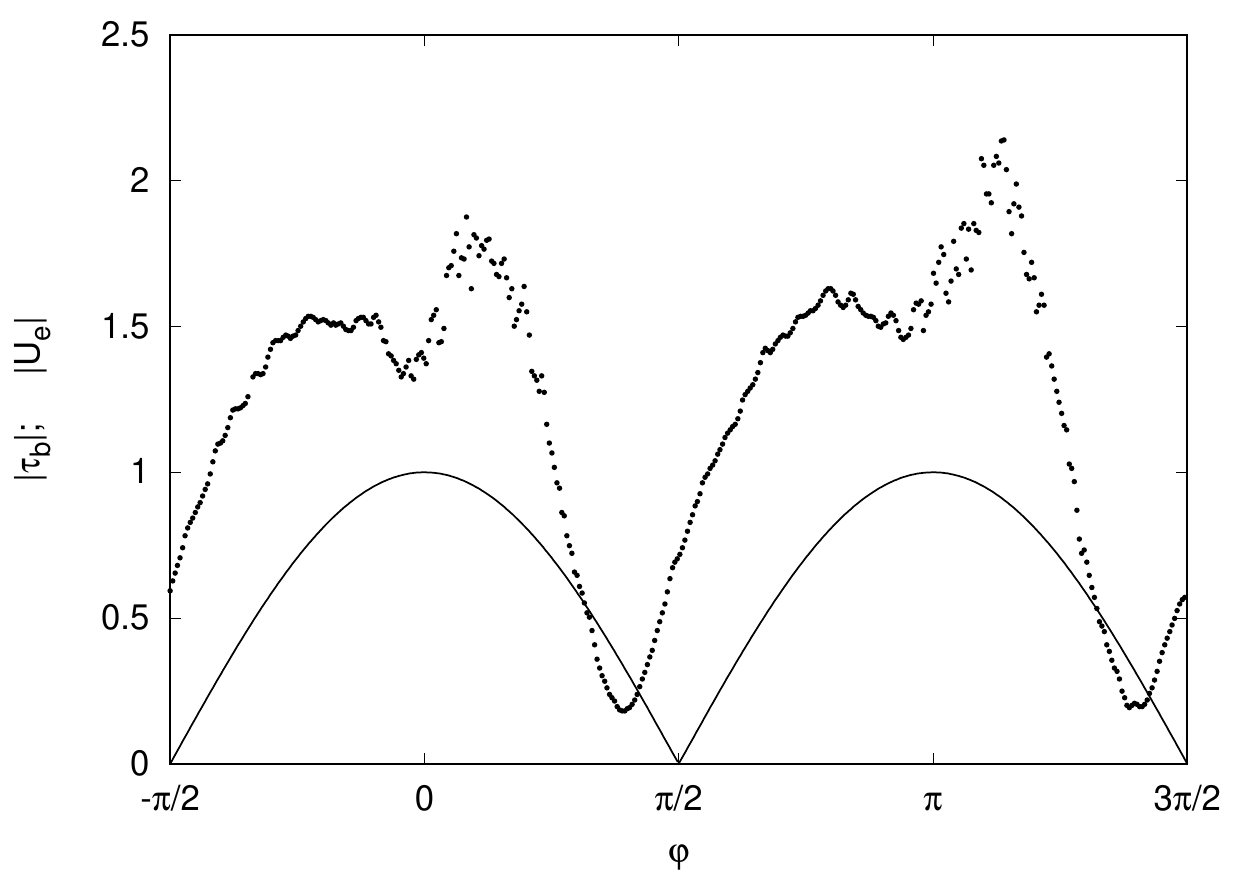}}
  \fi
\end{picture}
\caption{%
Dimensionless wall shear stress $|\tautot|=|\tautot^*|/\left(\frac{1}{2}\densf^* \U^* \om^* \del^*\right)$ plotted versus the phase $\phase$ of the cycle for a smooth wall and $\Rdel=761$. %
The dots are the experimental data by \citet{jensen1989} and the continuous line is the dimensionless external velocity magnitude in the experiment, $|U_e|=|U_e^*|/\U^*$. %
}%
\label{fig1}
\end{figure}

For small values of $\ds$, transition to turbulence occurs in the early stages of the decelerating phases, in a way apparently similar to that over a smooth-wall \citep{mazzuoli2016b}. %
The effect of the transition to turbulence can be observed in figure \ref{fig1}, where the bottom shear stress measured by \citet{jensen1989} at the bottom of an oscillatory boundary layer over a smooth bottom for $\Rdel=761$ is plotted versus the phase $\phase$ of the cycle. %
The phase variable, $\phase\in[0,2\pi[$, is expressed in radiants and defined in order to be equal to zero when the maximum absolute value of the velocity far from the bottom is attained. %
Turbulence appears during the decelerating phases, but the flow recovers a laminar-like behaviour during the accelerating phases (\textit{intermittently turbulent regime}). %
As discussed in \citet{blondeaux2018}, it can be easily verified that the intermittently turbulent regime is present in a significant part of the coastal region which shifts towards the shore during mild wave conditions while it shifts towards the offshore region during storms. %

Then, by increasing either $\Rdel$ or $\ds$, transition to turbulence occurs earlier and earlier, therefore pervading also the accelerating phase. %
It is noteworthy that, for $\Rdel>150$ and $\ds=6.95$, the wall is hydrodynamically rough and turbulent fluctuations practically never disappear \citep{mazzuoli2016b,ghodke2016} even though turbulence strength during the decelerating phases differs from that observed during the accelerating phases. %
In fact, one of the difficulties of modelling of the wave-averaged sediment transport induced by propagating surface waves lies in the fact that the sediment transport rate during the accelerating phases of the cycle differs from that observed during the decelerating phases even though the free stream velocity has the same value. %
Although in a large number of empirical formulae used to quantify the sediment transport rate, the sediment flux is independent of the sign of the flow acceleration, figure~\ref{fig1} suggests that the bottom shear stress and the sediment transport rate observed during the decelerating phases are associated to levels of turbulence much larger than those observed during the accelerating phases. %
%
%%
%%%% Figure 1  *****************************************************
% ******************************************************************
%%
A further difficulty comes from the experimental evidence that the threshold conditions for the initiation of sediment motion differ from those leading particles to stop. %
In the former case, the probability that sediment particles are set into motion depends on the occurrence of a favourable particle-flow interaction, while in the latter situation, particles can stop moving depending on the likelihood that they find a stable configuration on the bed surface \citep{clark2017}. %

The results of the DNSs, which are described in the following, are aimed at verifying that the picture we have drawn previously is realistic and significantly affects sediment dynamics. %
In particular, we want: (\emph{I}) to verify whether the differences in the hydrodynamics of the boundary layer and the dynamics of sediment grains during the accelerating and decelerating phases of the wave-induced bottom flows are significant; (\emph{II}) to evaluate the dependence of the sediment flow rate $\qpmean^*$ on quantities characterising the flow properties, like the bottom shear stress $\tau_b^*(t)$ or the turbulent kinetic energy, for values of the parameters such that the flow regime is intermittently turbulent (see table~\ref{tab0}). %

The paper is structured as follows. %
In the next section we formulate the problem and we briefly describe the numerical approach used to evaluate the flow field and the sediment dynamics. %
In section \ref{res}, we describe the flow field and the sediment transport. %
Section \ref{conc} is devoted to the conclusions. %
%

%*** Table 1 *******************************************************
\begin{table}
	\begin{center}
		\begin{tabular}{l r r r r r c r c r c}
		\hline
		\multirow{1}{.5cm}{Run} & 
		\multirow{1}{.5cm}{$\Rdel$} &
		\multirow{1}{.5cm}{$\Rd$} & 
		\multirow{1}{.5cm}{$\mob$} & 
		\multirow{1}{.5cm}{$Re_p$} & 
		\multirow{1}{.5cm}{$\Kc$} & 
		\multirow{1}{.5cm}{$s$} &
		\multirow{1}{.5cm}{$\dfrac{\del_{dis}^*}{\del^*}$} &
		\multirow{1}{1.1cm}{$\dfrac{u_{\tau,max}^*\ds^*}{\nu^*}$} &
		\multirow{1}{1.1cm}{$\max_t \vert\shields\vert$} &
		\multirow{1}{1cm}{\centering simulated\\ cycles}\vspace*{.4cm}\\ 
		\hline
		\runa & $450$  & $150.7$ & $11.12$  & $45.2$  & $672$  & $2.65$ & $1.41$ & $12$ & $0.07$ & $4$ \\
		\runb & $750$  & $251.3$ & $30.89$  & $45.2$  & $1119$ & $2.65$ & $4.85$ & $21$ & $0.22$ & $2$ \\
		\runbf & $750$ & $251.3$ & $30.89$  & $45.2$  & $1119$ & $2.65$ & $2.81$ & $18$ & $0.16$ & $2$ \\
		\rund & $1000$ & $335.0$ & $60.50$  & $43.1$  & $1493$ & $2.65$ & $7.15$ & $25$ & $0.33$ & $1$ \\
		\hline
		\end{tabular}
	\end{center}
\caption{%
Flow parameters for the present runs. From left to right: the Reynolds numbers $\Rdel=\U^*\del^*/\nu^*$ and $\Rd=\U^*\ds^*/\nu^*$, the mobility number $\mob=\U^{*2}/\vs^{*2}$, $\vs^*=\sqrt{(s-1)\g^*\ds^*}$ indicating the sediment fall velocity,	the Reynolds number of the sediment $Re_p=\vs^*\ds^*/\nu^*$ (also known as Galileo number), the Keulegan-Carpenter number $\Kc=\U^*/(\om^*\ds^*)$ and the specific gravity $s=\denss^*/\densf^*$. %
Note that the ratio $d^*/\del^*=\Rd/\Rdel$ is equal to $0.335$ for all the runs. %
The dimensionless displacement thickness $\del_{dis}^*/\del^*$, the grain Reynolds number $u_{\tau,max}^*\ds^*/\nu^*$ and the maximum value of the Shields parameter $\vert\theta\vert$ are also shown, with the maximum friction velocity defined as $u_{\tau,max}^*=\max_t \sqrt{\vert\tau_b^*\vert/\densf^*}$. %
The last column gives the number of periods used in the post-processing, after the transient was removed. %
}%
\label{tab0}
\end{table}
%*******************************************************************

\section{Formulation of the problem and numerical approach}
\label{problem}

The flow within the boundary layer at the bottom of a sea wave is investigated by assuming that the wave steepness, i.e. the ratio between the amplitude and the length of the wave, is small and the linear Stokes theory describes the flow generated far from the bottom by wave propagation. %
Even though this approach neglects nonlinear effects and in particular the existence of a steady streaming, it provides a fair description of the oscillatory flow generated by propagating sea waves close to the bottom when their amplitude is small. %
Then, the flow within the bottom boundary layer can be determined by approximating it as the flow generated by an oscillating pressure gradient close to a fixed wall. %
Nonlinear effects, which become significant when the wave propagates into shallow waters because of the increase of its amplitude, are neglected. %
In particular, the presence of steady streamings and wave asymmetries, which produce a skewness of the flow velocity and acceleration in the OBL \citep{vandera2011,scandura2016}, are not presently considered. %
Hence the pressure gradient, that drives the flow, can be written in the form: %
\begin{equation}
\frac{\partial p^*}{\partial \xf{1}^*} 
= 
- \densf^* \U^* \om^*\sin{(\om^*t^*)}; 
\qquad 
\frac{\partial p^*}{\partial \xf{2}^*} 
= 
0; 
\qquad 
\frac{\partial p^*}{\partial \xf{3}^*} 
= 
0
\label{pres}
\end{equation}
where $\xff{*}$ is a Cartesian coordinate system such that $\xf{1}^*$-axis points in the direction of wave propagation and the $\xf{2}^*$-axis is vertical and points in the upward direction. %
The pressure gradient described by \eqref{pres} drives the fluid motion as well as the motion of spherical particles of density $\denss^*$ and diameter $\ds^*$ which mimic actual sediment grains. %
The initial position of the spheres is obtained by simulating the settling of a large number $\np$ of particles in the still fluid until the particles accumulate on the plane $\xf{2}^*=0$. %
Then, the particles in contact with the plane $\xf{2}^*=0$ are kept fixed while the others are free to move. %
The thickness $x_{2\,bottom}^{(init)}$ of the particle layer, at the beginning of each run, is indicated in table~\ref{tab1}. %
However, the reader should be aware that the fluid action is able to move only a few surficial layers of particles and many layers of particles practically do not move during the simulations. %
%
%*** Table 2 *******************************************************
\begin{table}
	\begin{center}
		\begin{tabular}{l r r r c r r c c}
		\hline
		\multirow{1}{*}{Run} & 
		\multirow{1}{*}{$\Lx$} &
		\multirow{1}{*}{$\Ly$} & 
		\multirow{1}{*}{$\Lz$} & 
		\multirow{1}{*}{$x_{2\,bottom}^{(init)}$} & 
		\multirow{1}{*}{$\nx$} & 
		\multirow{1}{*}{$\ny$} & 
		\multirow{1}{*}{$\nz$} &
		\multirow{1}{*}{$\np$}\vspace*{.12cm}\\ 
		\hline
		\runa & $25.73$ & $30.01$ & $12.86$ & $6.6$ & $768$  & $896$  & $384$ & $50503$ \\
		\runb & $24.50$ & $30.63$ & $12.25$ & $7.6$ & $1024$ & $1280$ & $512$ & $50557$ \\
		\rund & $24.50$ & $36.75$ & $12.25$ & $6.8$ & $1024$ & $1536$ & $512$ & $61552$ \\
		\hline
		\end{tabular}
	\end{center}
\caption{%
Size of the computational domain, initial bed elevation, number of grid points and of sediment particles. %
}%
\label{tab1}
\end{table}
%*******************************************************************

\subsection{The fluid motion}

The hydrodynamic problem is written in dimensionless form introducing the following variables: %
\begin{align}
t 
= 
t^* \om^*; 
\qquad 
\xff{}
= 
\frac{\xff{*}}{\del^*}; 
%\qquad 
\\
\uff{} 
= 
\frac{\uff{*}}{\U^*}; 
\qquad 
p 
= 
\frac{p^*}{\densf^*\U^{*2}} \:\:.
%\]
\label{var}
\end{align}
In (\ref{var}), $t^* $ is time and $\uf{1}^*, \uf{2}^*, \uf{3}^*$ are the fluid velocity components along the $\xf{1}^*$-, $\xf{2}^*$- and $\xf{3}^*$-directions, respectively. %

Using (\ref{var}), continuity and momentum equations read: %
\begin{equation}
\label{equ2}
\frac{\partial u_j }{\partial x_j} 
= 
0 
\end{equation}
\begin{equation}
\label{equ1}
\frac{\partial u_i}{\partial t} 
+ 
\frac{\Rdel}{2} u_j \frac{\partial u_i}{\partial x_j} 
= 
- 
\frac{\Rdel}{2}\frac{\partial p}{\partial x_i} 
+ 
\delta_{i1} \sin{(t)} 
+ 
\frac{1}{2}\frac{\partial^2 u_i }{\partial x_k \partial x_k} 
+ 
f_i
\end{equation}
where the pressure gradient is written as the sum of two terms. %
One term ($-\sin{(t)}$) is the imposed streamwise pressure gradient that is uniform and drives the fluid oscillations. %
The other term ($\frac{\Rdel}{2} \frac{\partial p}{\partial x_i}$) is associated with the vortex structures shed by the sediment grains or with the turbulent eddies and is an output of the numerical simulations. %

At the lower boundary of the fluid domain ($\xf{2}=0$), where a rigid wall is located, the no-slip condition is enforced: %
\begin{equation}
\uff{} = (0,0,0) \:\:,
%\qquad 
%\mbox{at} 
%\qquad 
%\xf{2} = 0
\label{bc1}
\end{equation}
while at the upper boundary ($\xf{2}=\Ly$) the free stream (free slip) condition is enforced: %
\begin{equation}
\left( 
\dudx{\uf{1}}{2}{}
,\,
\dudx{\uf{3}}{2}{} 
\right) 
= 
(0,0)\, ; 
\qquad 
\uf{2} = 0 \ \ .
%\qquad \mbox{at} \qquad \xf{2} = \Ly
\label{bc1x}
\end{equation}
Moreover, periodic boundary conditions are enforced in the homogeneous directions ($\xf{1}, \xf{3}$), because the computational box is chosen large enough to include the largest vortex structures of the flow. %

The hydrodynamic problem is solved numerically by means of a finite difference approach in a computational domain of dimensions $\Lx, \Ly$ and $\Lz$ in the streamwise, wall-normal and spanwise directions, respectively. %
A uniform grid is introduced with $\nx,\,\ny,\,\nz$ grid points along the three directions. %

The numerical scheme is the same as that used by \citet{Aman2014a,mazzuoli2016a,kidanemariam2017,mazzuoli2019}. %
Standard centrered second-order finite difference approximations are used to approximate the spatial derivatives, written using a uniform, staggered Cartesian grid, while the time-advancement of Navier-Stokes equations is made using a fractional-step method based upon the combination of explicit (three-step Runge-Kutta) and implicit (Crank-Nicolson) discretisations of the nonlinear and viscous terms, respectively. %

The continuity and momentum equations are solved throughout the whole computational domain including the space occupied by the solid particles, which are immersed in the fluid and move close to the bottom. %
The no-slip condition at the sediment-fluid interface is enforced, using the immersed-boundary technique \citep{Uhlmann2005}, by means of the terms $f_i$, added to the right hand side of (\ref{equ1}). % 
The numerical code has been widely tested \citep[see for example][]{mazzuoli2016a}. %

\subsection{The sediment motion}

The sediment grains, which are modelled as spherical particles of uniform diameter $d^*$, are moved according to Newton-Euler equations: %
\begin{align}
%\mpart^* \frac{d {\bf u}_p^*}{dt^*} 
%&= 
%\int_{S^*} {\bf T}^*\cdot {\bf n} dS^* + {\bf W}^* + {\bf F}^*_p\\
%%
%I^*_p \frac{d {\bf \omega}_p^*}{dt^*} 
%&= 
%\int_{S^*} {\bf r}^* \times \left( {\bf T}^*\cdot {\bf n} \right) dS^* +{\bf T}_p^*
%%
\mpart^* \frac{d \up{i}}{dt^*} 
&= 
\int_{S^*} \stenf \, n_j dS^* + W_i^* + F_i^{(p)*}\\
I^*_p \frac{d \avel{i}}{dt^*} 
&= 
\int_{S^*} \epsilon_{ijk}\ r_j^* \, \sigma_{km}^{(f)*} \, n_m dS^* + T_i^{(p)*}
\end{align}
where $m^*_p$ is the mass of a single spherical particle, $I^*_p$ is its moment of inertia and $\epsilon_{ijk}$ denotes the Levi-Civita symbol. %
Moreover, $\up{i}$ and $\avel{i}$ are the $i$-th components of the particle linear and angular velocity, respectively ($i=1,\,2,\,3$). %
Finally, $\stenf$ is the fluid stress tensor, $r_j^*\ (j=1,\,2,\,3)$ the vector from the centre of the particle to the generic point on its surface, $n_m\ (m=1,\,2,\,3)$ is a normal unit vector pointing outward from the surface of the particle, $W_i^*$ is the weight of the particle and $F_i^{(p)*}$ and $T_i^{(p)*}$ indicate the force and torque due to inter-particle collisions. %
It follows that the phenomena associated with the grain size distribution and the irregular shape of the sand grains are not considered. %

The motion of the sediment grains turns out to be controlled by their specific gravity $s=\denss^*/\densf^*$ and their dimensionless size $\ds=\ds^*/\del^*$ even though it is common to use also the particle Reynolds number $R_p=\sqrt{(s-1)\g^*\ds^{*3}}/\nu^*$ (often known also as Galileo number). %
The values of the parameters for the simulations presently considered are indicated in table~\ref{tab0} while the size of the computational domain and the number of grid points employed in each run are listed in table~\ref{tab1}. %
In particular, one DNS was carried out, for the same values of the parameters as those of $\runb$, by fixing the spheres at their resting position. %
This run is indicated by ``$\runbf$'' in table~\ref{tab0}. %

The force and torque due to the grain-grain contacts are evaluated by means of a discrete-element model (DEM) which is based upon a linear mass-spring-damper model of particle interaction. %
More details on the evaluation of particle dynamics can be found in \citet{Aman2014b}. %

Since the temporal scale of the grain collisions is ${\cal O}(100)$ times smaller than the temporal scale of the oscillating flow, the position of colliding particles is evaluated by splitting each time step of the fluid solver into ${\cal O}(100)$ substeps, during which the hydrodynamic force is assumed to be constant. %
The DEM model asks for the specification of the values of the following parameters: the ``force range'', the normal stiffness, the Coulomb friction coefficient and the value of the restitution coefficient. %
These parameters are given values essentially equal to those of \citet{mazzuoli2016a} (see table~\ref{tab3}). %
%
%*** Table 3 *******************************************************
\begin{center}
\begin{table}
\begin{center}
    \begin{tabular}{c r l c c c}
    \hline
  \multirow{2}{*}{Run}   & \multirow{2}{*}{$k_n$} & \multirow{2}{*}{$\mu_{cf}$} & \multirow{2}{*}{$\varepsilon_d^{(part)}$}  & \multirow{2}{*}{$\delta^*_c/\Delta x^*$} & \multirow{2}{*}{$\ds^*/\Delta x^*$} \vspace*{.4cm}\\ \hline
     \runa & $711.2$  & $0.4$ & $0.9$ & $1.0$ & $10$ \\
     \runb & $1439.2$ & $0.4$ & $0.9$ & $1.0$ & $14$ \\
     \rund & $3384.0$ & $0.4$ & $0.9$ & $1.0$ & $14$ \\
    \hline
    \end{tabular}
\end{center}
\caption{%
Values of the DEM parameters: dimensionless normal stiffness $k_n=\frac{6}{\pi}\frac{k_n^*~\Delta x^*}{\ds^{*3}\g^*\denss^*}$, Coulomb friction coefficient $\mu_{cf}$, restitution coefficient $\varepsilon_d^{(part)}$ and the ``force range'' $\delta^*_c/\Delta x^*$. %
}%
\label{tab3}
\end{table}
\end{center}

\subsection{Average operators}
Since the bed surface preserves essentially a horizontal profile during each phase of the oscillation period for all the simulations presently considered, both the flow and the particle motion are assumed statistically homogeneous over horizontal planes. %
Thus plane averages are performed of quantities associated with either fluid or particle phases using the definitions provided by \citet{Aman2014b}. %
In particular, with reference to the sample box $\vot(\xb{})$ of size $\Lx\times\dy\times\Lz$, centered in $\xb{}$, the plane average, $\sav{\varf}$, of the generic fluid property $\varf$ is computed as: %
\begin{equation}
\sav{\varf}(\xb{},t) 
= 
\dfrac{% 
\Int_\vot\varf(\xb{},t)\,\Sump\idff(\xb{}-\xb{\iind}(t))~dV%
}{%
\Int_\vot\Sump\idff(\xb{}-\xb{\iind}(t))~dV\:\:,%
}%
\label{eq:avf}
\end{equation}
while the plane average of a particle property, say $\var$, is defined as: %
\begin{equation}
\sav{\var}(\xb{},t) 
= 
\dfrac{% 
\Int_\vot\var(\xb{},t)\,\Sump\idf (\xb{}-\xb{\iind}(t))~dV%
}{%
\Int_\vot\Sump\idf (\xb{}-\xb{\iind}(t))~dV%
}%
\:\:,
\label{eq:avp}
\end{equation}
where $\xb{\iind}(t)$ is the position of the centre of the generic $\ell$-particle. %
Moreover, the particle indicator function $\idf(\xb{}-\xb{\iind}(t)) $ is equal to $1$ if $\vert\xb{}-\xb{\iind}(t)\vert < d/2$ or to $0$ otherwise, while $\idff(\xb{}-\xb{\iind}(t))=1-\idf$. %
In the following, where not explicitly indicated, plane-average quantities are shown omitting the angular brackets for the sake of clarity. %

\section{Discussion of the results}
\label{res}

When the bottom is made up of moving sediment grains, the DNS of Navier-Stokes and continuity equations within the bottom boundary layer generated by an oscillatory pressure gradient requires huge computational resources and ``wall clock'' time. %
Hence, only a few cases are considered in the following and attention is focused on values of the parameters such that sediment particles are set into motion and the Reynolds number is large enough to trigger turbulence appearance. %
The values of the relevant dimensionless parameters of the present simulations are indicated in table~\ref{tab0}. %
At this stage it is worth pointing out that neither ripples nor transverse bands of sediments, like those detected by \citet{blondeaux2016}, appear during the simulations. %
This is an effect of the size of the domains, which is too small for ripples to appear, and of the duration of the simulations, which is too short for ripples to develop. %
Indeed, we wanted to consider the plane bottom configuration. %

For illustration, consider $\runb$ and $\rund$ which are characterised by the dimensionless diameter $\ds$ equal to $0.335$ and the Reynolds number $R_\delta$ equal to $750$ and $1000$, respectively. %
In order to relate these values of $R_\delta$ and $\ds$ to a field case, it can be easily verified that $R_\delta=750$ is the Reynolds number of the bottom boundary layer generated by a surface wave characterised by a period $T^*$ equal to $7$~s and a height $H^*$ equal to about $1.4$~m propagating 
in a coastal region where the water depth $h^*$ is equal to $10$~m. %
It turns out that $\del^*=1.49$~mm and that $\ds=0.335$ implies $\ds^*=0.5$~mm, i.e. a grain size which is coincident with the limit between medium and coarse sand. %
\smallskip
\subsection{Appearance of turbulence and turbulent kinetic energy}
As discussed in \citet{sleath1988}, on the basis of \citet{kajiura1968}'s criterion, turbulence appearance is certainly triggered during the oscillatory cycle for such values of the parameters ($\ds=0.335,\,R_\delta=750$). %
Indeed, the condition suggested by \citet{kajiura1968} for the initiation of the turbulent regime, i.e. $\U^*\ds^*/\nu^* \ge 104$, is widely satisfied. %
It is worth pointing out that, using the present notation, \citet{kajiura1968}'s criterion can be written in the form $\Rdel \ge 104 /\ds$. %
However, laboratory measurements \citep[e.g.][]{sleath1988} suggest that the random fluctuations of the velocity are not large and appear only during a part of the oscillatory cycle. %
%%
%%%% Figure Sketch *****************************************************
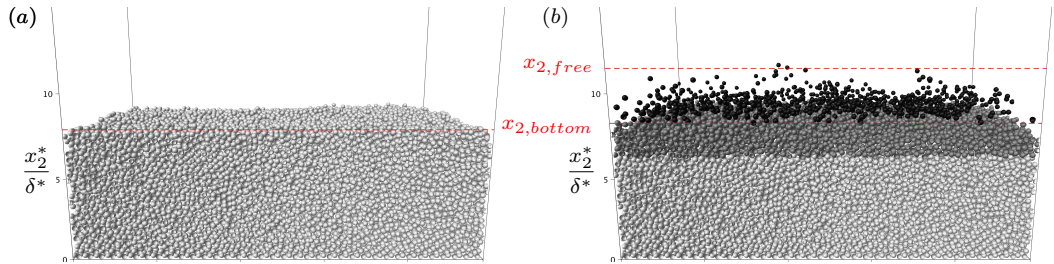
\begin{figure}
\begin{picture}(0,100)(0,0)
  \ifsix
  \put(5,0){\includegraphics[trim=0cm 0cm 0cm 0cm, clip, width=.50\textwidth]{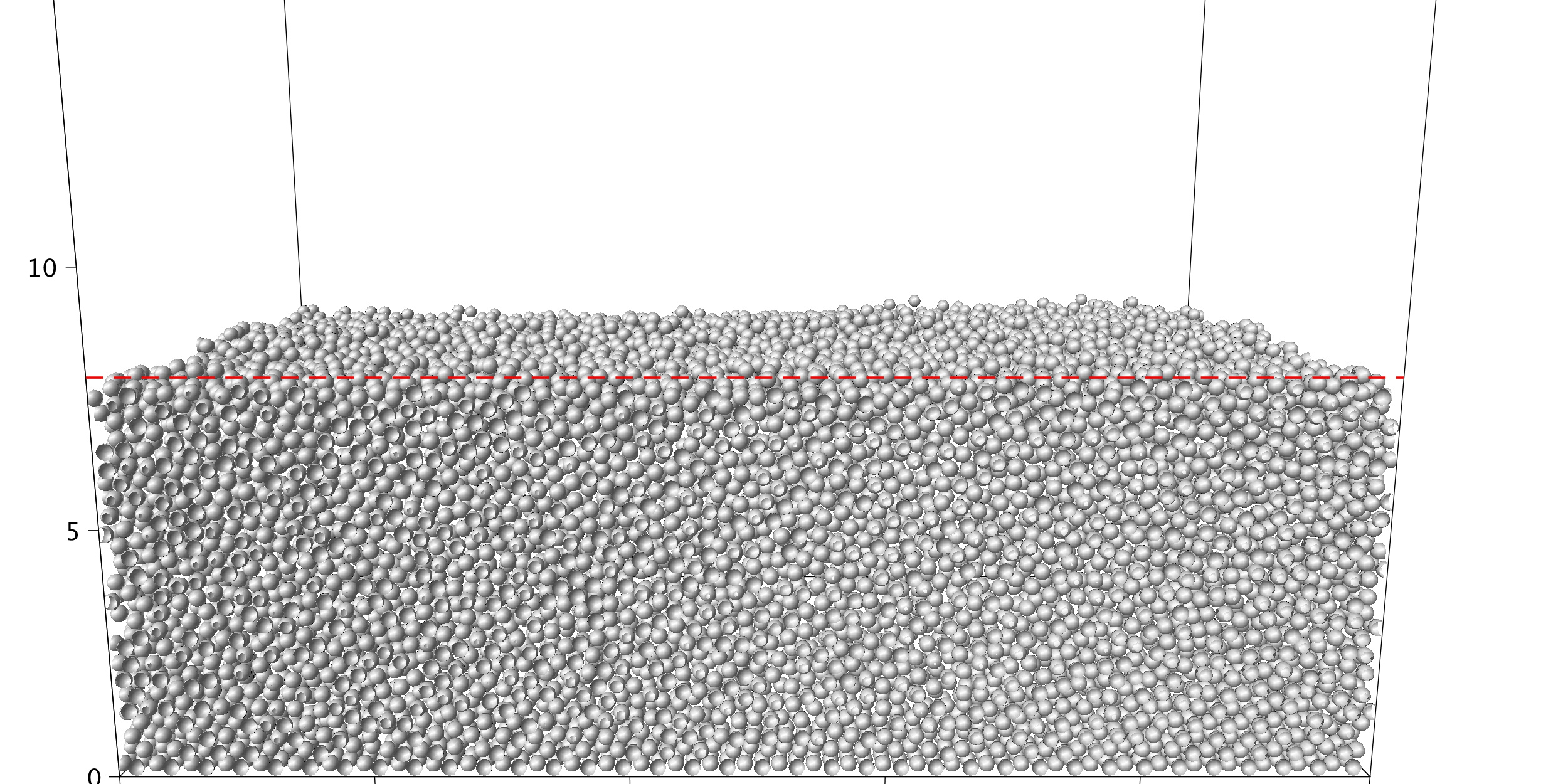}}
  \put(210,0){\includegraphics[trim=0cm 0cm 0cm 0cm, clip, width=.50\textwidth]{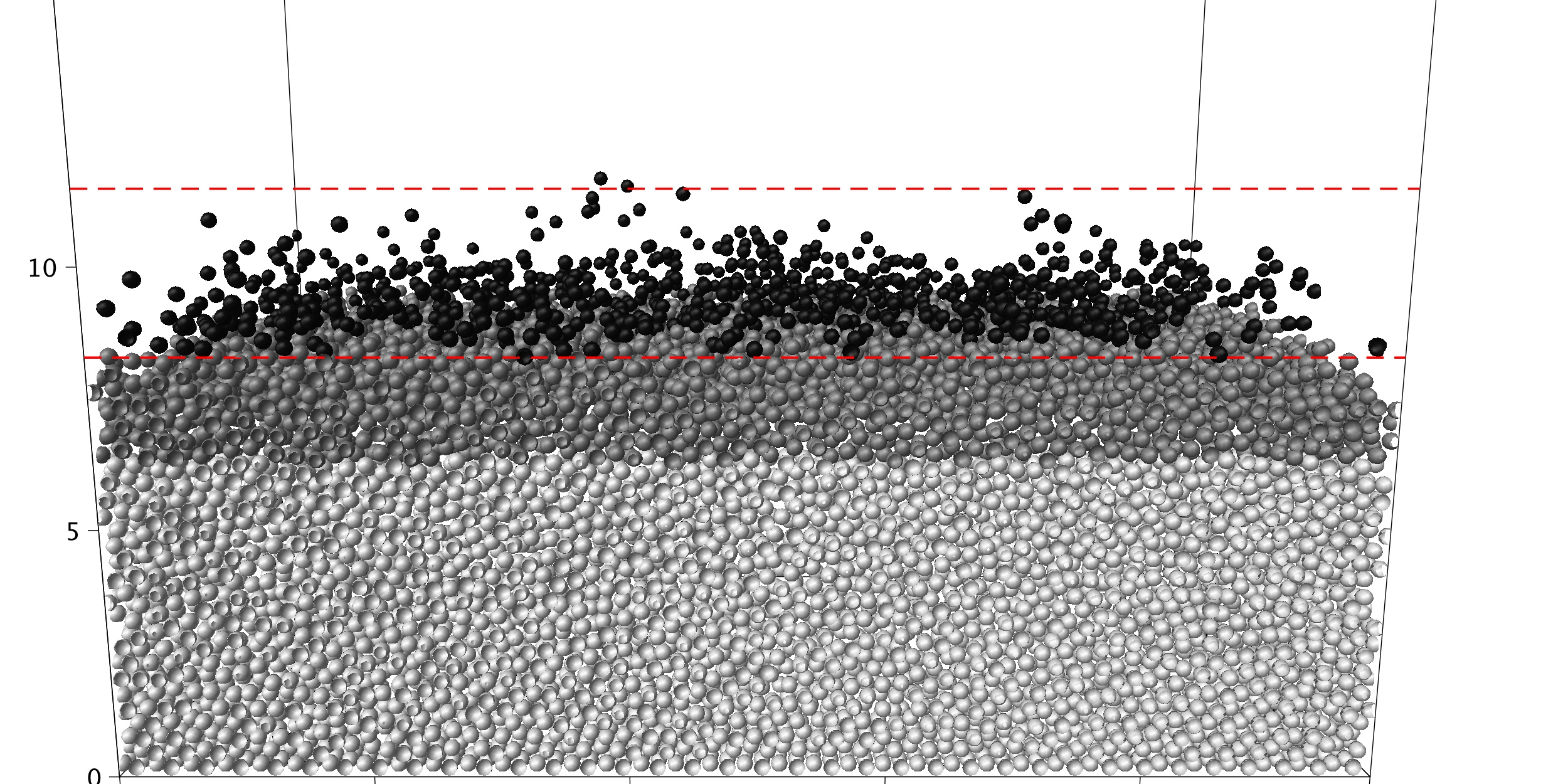}}
%  \put(-9,50){\footnotesize \color{red!}$x_{2,bottom}$}
  \put(180,50){\footnotesize \color{red!}$x_{2,bottom}$}
  \put(188,73){\footnotesize \color{red!}$x_{2,free}$}
  \put(-5,90){\footnotesize $(a)$}
  \put(195,90){\footnotesize $(b)$}
  \put(-5,90){\footnotesize $(a)$}
  \put(204,32){\footnotesize $\dfrac{\xf{2}^*}{\del^*}$}
  \put(0,32){\footnotesize $\dfrac{\xf{2}^*}{\del^*}$}
  \fi
\end{picture}
\caption{%
Bed configuration $(a)$ at the initial state and $(b)$ at the early deceleration phase $\phase=0.02$ ($\rund$). %
The broken horizontal lines indicate the bottom surface elevation and the maximum elevation reached by particles. %
Light gray particles are resting while the other particles are moving and the black ones lay above $x_{2,bottom}$. %
}%
\label{figSketch}
\end{figure}
% ******************************************************************

The flow and sediment dynamics were simulated within a computational box $24.50~\delta^*$ long, $12.25~\delta^*$ wide and $30.63~\delta^*$ high (see table \ref{tab1} summarizing the main parameters of the numerical box and grid). %
The size of the box is similar to that used by \citet{Verzicco1996} and \citet{Vittori1998} for their simulations of turbulence dynamics in an oscillatory boundary layer and turns out to be large enough for turbulence generation (minimal flow unit). %
Only the height of the box is significantly larger because, in the present simulation, a large number of spherical particles deposited on the bottom. %
Before starting each simulation, particles were located randomly over the computational domain and let them settle in resting fluid. %
Then, they were ``shaken' until a closely-packed configuration was attained. %
Finally, the layer of particles in contact with the wall was fixed while those the center of which was located above the elevation $x_{2\,bottom}^{*(init)}$, indicated in table~\ref{tab1}, were removed to guarantee the plane bottom configuration (cf. figure~\ref{figSketch}a). %
The number of the remaining particles is indicated in table~\ref{tab1}. %
The grid size is equispaced and uniform along the three coordinates and such that $10$ grid points are present per grain diameter. %
%
%%%% Figure 2  *****************************************************
\begin{figure}
\begin{picture}(0,260)(0,0)
  \iftwo
  \put(0,0){\includegraphics[trim=0cm 22.2cm 0cm 0cm, clip, width=.47\textwidth]{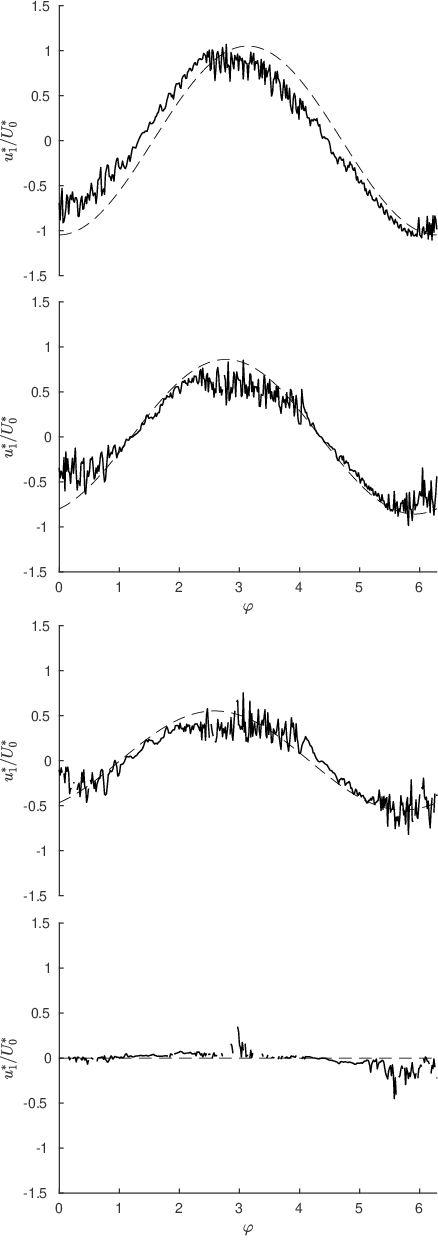}}
  \put(200,0){\includegraphics[trim=0cm .2cm 0cm 21.7cm, clip, width=.47\textwidth]{\figdir figure3}}
  \put(30,245){\footnotesize $(a)$}
  \put(30,122){\footnotesize $(b)$}
  \put(230,245){\footnotesize $(c)$}
  \put(230,122){\footnotesize $(d)$}
  \fi
\end{picture}
\caption{%
Streamwise velocity component plotted versus the phase $\varphi$ during the second cycle for $\xf{1}=\Lx/2$ and $\xf{3}=\Lz/2$ and different values of $\xf{2}$. %
$(a)$ $\xf{2}^*=x^*_{2,bottom}+3~\delta^*$, $(b)$ $\xf{2}^*=x^*_{2,bottom}+\delta^*$, $(c)$ $\xf{2}^*=x^*_{2,bottom}+0.5~\delta^*$, $(d)$ $\xf{2}^*=x^*_{2,bottom}$. %
Continuous line = numerical results; broken line = Stokes' solution. %
$\Rdel=750$, $\ds=\frac{\ds^*}{\del^*}=0.335$ ($\runb$). %
}%
\label{fig2}
\end{figure}
% ******************************************************************
 
To give a qualitative idea of the flow field that is generated close to the bottom for these values of the parameters and to show how turbulence appearance and the moving grains affect the velocity field, figure \ref{fig2} shows the streamwise velocity component plotted versus the phase $\phase$ during the second flow cycle, for different values of the distance $\xf{2}-x_{2,bottom}$ from the bottom and for $\xf{1}=\Lx/2$ and $\xf{3}=\Lz/2$. %
In the same figure, the Stokes solution is also plotted to allow an easy comparison of the numerical results with the laminar solution. %
The distance of the numerical velocity probes from the time-average bottom elevation $x_{2,bottom}$ is evaluated assuming that the instantaneous bottom surface elevation coincides with the plane where the plane-averaged volume fraction of the solid particles $\sav{\svf}$ reaches the value $0.1$ \citep[see][for further details on the computation of the bed surface]{mazzuoli2019}. %
It turns out that $x^*_{2,bottom}=6.60~\del^*$ with fluctuations of the bottom elevation, during the flow cycle, ranging between $6.51~\del^*$ and $6.80~\del^*$ (i.e. of the order of $\ds^*$). %
This heuristic assumption might be modified taking into account that significant local instantaneous fluid velocities can be found even for $\xf{2}$ smaller than $x_{2,bottom}$ when the Reynolds number is large enough to induce sediment motion and the spherical particles start to slide, roll and saltate on the resting particles. %
For example, the bottom position might be determined either by choosing a different threshold value of $\svf$ or by evaluating the value of $\xf{2}$ at which either the average streamwise velocity component or the turbulent kinetic energy vanish. %
Figure~\ref{figSketch}b shows that saltating and floating particles can be present above $x_{2,bottom}$ up to the level $x_{2,free}$ which delimits the particle free region. %

\begin{figure}[!]
\begin{picture}(0,300)(0,0)
  \iftwo
  \put(0,-10){
  \put(10,152){\includegraphics[trim=2cm 6cm 3cm 5cm, clip, width=.4\textwidth]{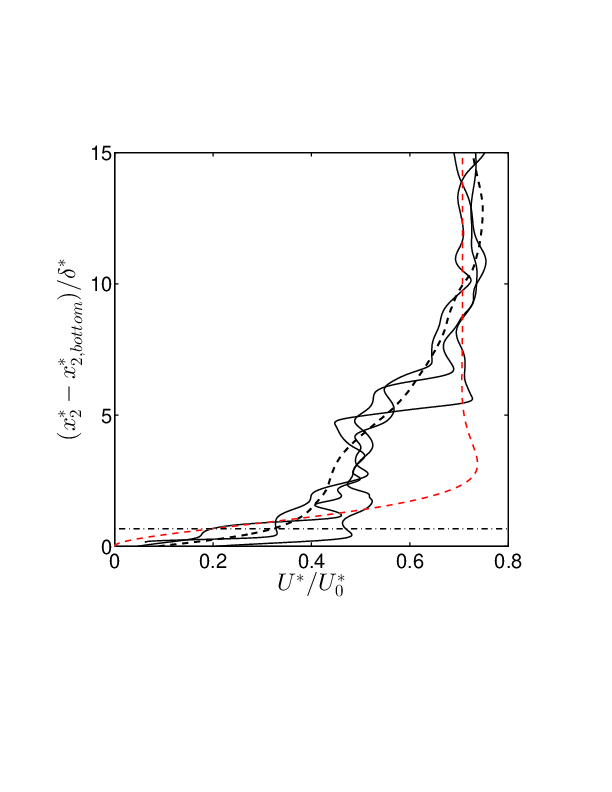}}
  \put(190,152){\includegraphics[trim=2cm 6cm 3cm 5cm, clip, width=.4\textwidth]{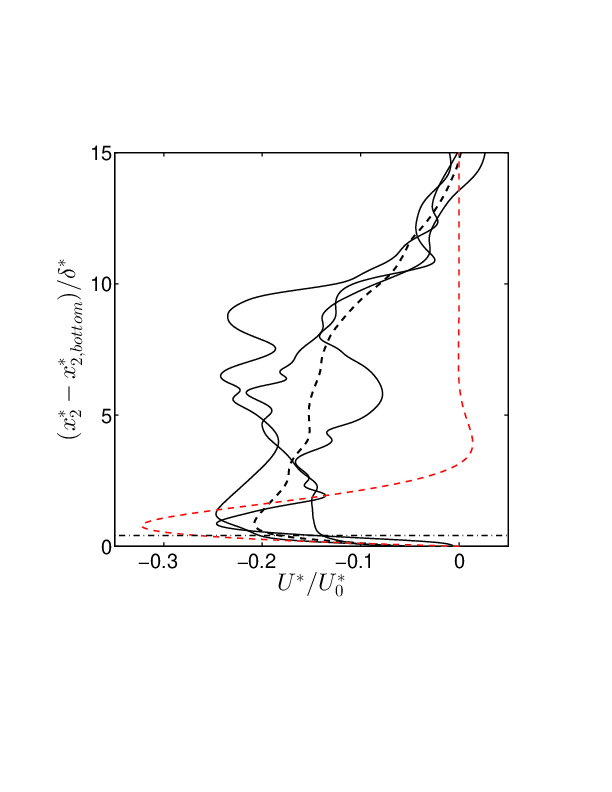}}
  \put(10,-2){\includegraphics[trim=2cm 6cm 3cm 5cm, clip, width=.4\textwidth]{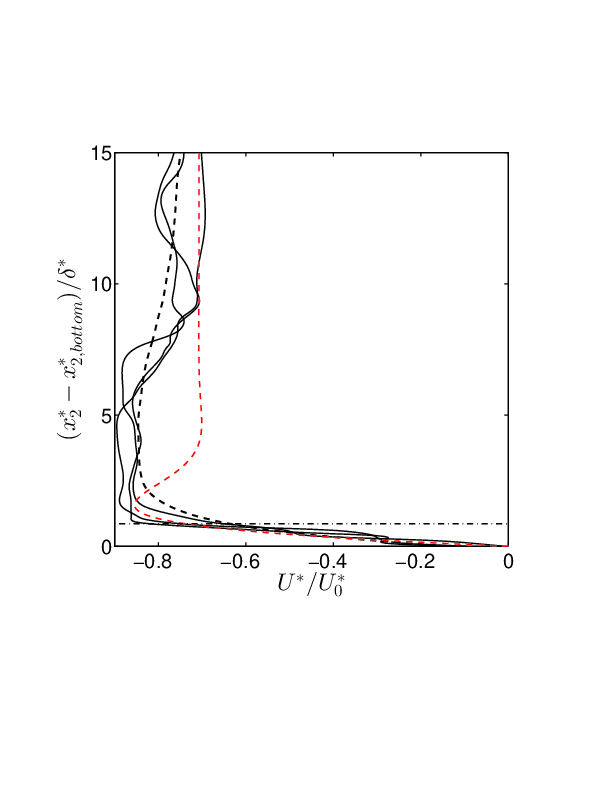}}
  \put(190,-2){\includegraphics[trim=2cm 6cm 3cm 5cm, clip, width=.4\textwidth]{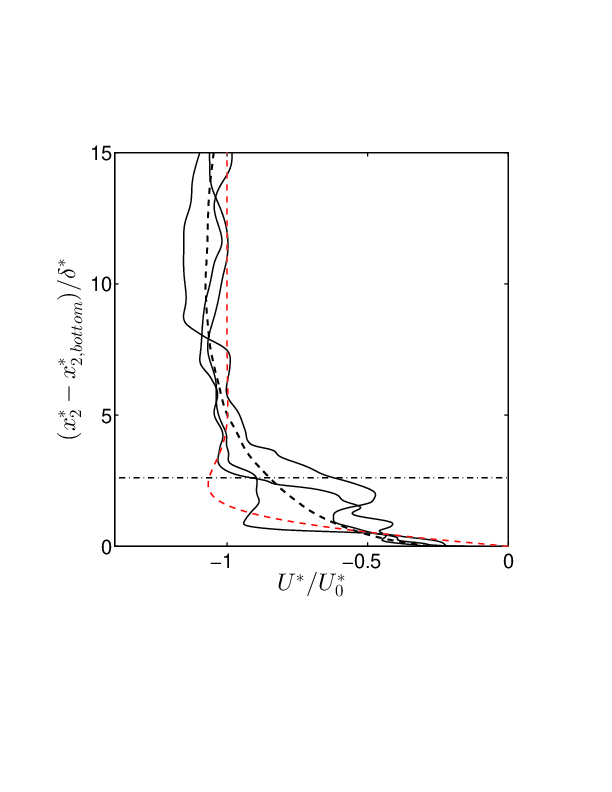}}
  }
  %\put(100,-3){\scriptsize $\om^*\phase^*$}
  \put(4,285){\footnotesize $(a)$}
  \put(180,285){\footnotesize $(b)$}
  \put(4,134){\footnotesize $(c)$}
  \put(180,134){\footnotesize $(d)$}
  \put(44,278){\footnotesize  $\phase = 1.25\,\pi$}
  \put(220,278){\footnotesize $\phase = 1.50\,\pi$}
  \put(104,127){\footnotesize $\phase = 1.75\,\pi$}
  \put(280,127){\footnotesize $\phase = 2.00\,\pi$}
  \fi
\end{picture}
\caption{%
Streamwise velocity component at $\xf{1}=\Lx/2$ and $\xf{3}=2,\,6,\,8$, plotted versus the vertical coordinate $\xf{2}-x_{2,bottom}$ in the near-bottom region. %
$(a)$ $\phase=1.25\,\pi$, $(b)$ $\phase=1.50\,\pi$, $(c)$ $\phase=1.75\,\pi$ and $(d)$ $\phase=2.00\,\pi$, $\phase$ being the phase during the second cycle (continuous line = local streamwise velocity; black broken line = plane-averaged streamwise velocity; red broken line = Stokes' solution). The horizontal dash-dotted line indicates the elevation $x_{2,free}$ (cf. figure~\ref{figSketch}b) above which the flow is free of particles. %
$\Rdel=750$, $\ds=0.335$ ($\runb$). %
}%
\label{figX}
\end{figure}
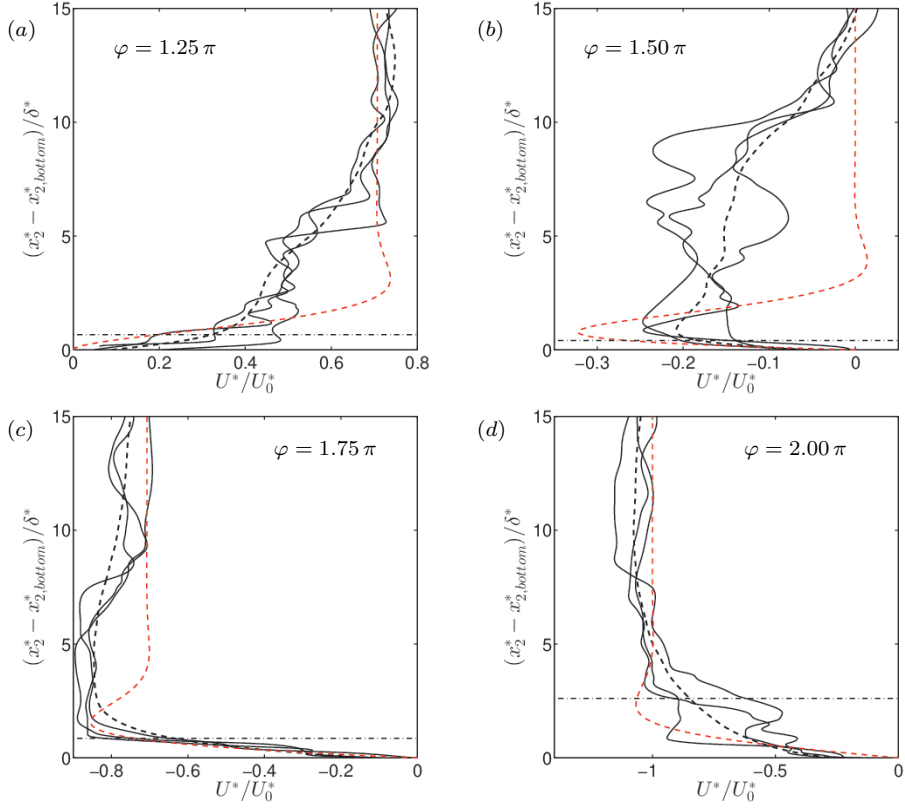

Figure~\ref{fig2} clearly shows that the velocity provided by the numerical simulation is characterised by large random fluctuations which appear when the velocity attains its largest values and at the beginning of the decelerating phases. %
However, these random velocity fluctuations are present only close to the bottom and they decrease moving far from it, till they assume negligible values when $\xf{2}-x_{2,bottom}$ is larger than about $15$ (not shown herein) where the velocity practically equals the free stream velocity. %
In particular, figure \ref{fig2}a and figure \ref{figX}, where the instantaneous velocity profile at $\xf{1}=\Lx/2$ is plotted versus $\xf{2}$ at different phases $\phase$ during the second cycle and for three different values of $\xf{3}$, show that the momentum transfer induced by turbulent fluctuations moves the overshooting of the velocity farther from the bottom and modifies its phase (cf. figure~\ref{figX}b). %
Consequently, the displacement thickness of the boundary layer, defined as %
\begin{equation}
\delta^*_{dis} 
=
\max_t 
\displaystyle\int_{x_{2\,bottom}^*}^{\Ly^*}
\left( 1-\dfrac{U^*}{\Ue^*}\right)\,dx_2^*
\:\:,
\label{eqdisp}
\end{equation}
is $4.85$ times larger than the displacement thickness computed for the Stokes' boundary layer, which is constant and equal to $\delta^*$ (in \eqref{eqdisp} $\Ue^*$ indicates the free-stream velocity). %
Such large increase of $\delta^*_{dis}$ is due to the presence of the particles saltating up to $\sim 8~\ds^*$ (i.e. $2.7~\del^*$) above the bed surface during the phases characterized by the maximum velocity (see the horizontal broken lines in figure~\ref{figX}d) which, in turn, enhance turbulent fluctuations far from the bottom. %
In fact, for $\runbf$ ($\Rdel=750$, $\ds=0.335$ and the spheres fixed at their resting positions), $\delta^*_{dis}$ is equal to $2.81~\delta^*$ (see also the values in table~\ref{tab0}). %
%
%%%% Figure X  *****************************************************
% ******************************************************************

\begin{figure}[!]
\begin{picture}(0,150)(0,0)
  \ifsix
  \put(0,0){\includegraphics[trim=0cm 0cm 0cm 0cm, clip, width=.50\textwidth]{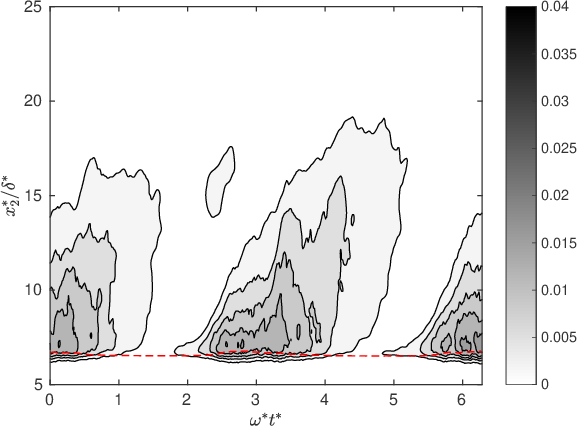}}
  \put(200,0){\includegraphics[trim=0cm 0cm 0cm 0cm, clip, width=.50\textwidth]{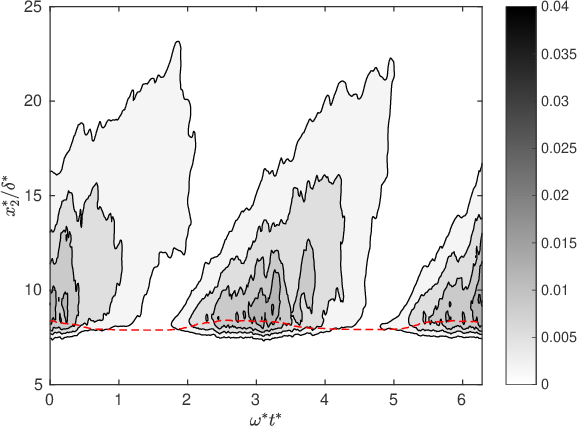}}
  \put(-5,130){\footnotesize $(a)$}
  \put(195,130){\footnotesize $(b)$}
  \fi
\end{picture}
\caption{%
Turbulent kinetic energy per unit volume, normalised with $\densf^*\U^{*2}$, plotted versus the phase $\phase$ during the second cycle for $\ds = 0.335$, $(a)$ $\Rdel = 750$ ($\runb$) and $(b)$ $\Rdel = 1000$ ($\rund$). %
The red broken line indicates the instantaneous bottom surface elevation. %
}%
\label{figY}
\end{figure}

The intermittent appearance of turbulence and its vanishing value far from the bottom clearly appear in figure~\ref{figY} where the dimensionless turbulent kinetic energy is plotted versus $\xf{2}$ and time during the second cycle of $\runb$ and $\rund$. %
The large computational costs do not allow a large number of oscillation cycles to be simulated and the turbulent kinetic energy is evaluated with respect to an average flow field that is not the phase averaged value but the plane-average of the velocity. %
This procedure makes the value of the turbulent kinetic energy at time $t$ slightly different from that at time $t+\pi$ and it makes the contour lines appearing in figure \ref{figY} not smooth because of the finite size of the computational domain. %
As already pointed out, turbulence is generated when the external velocity is maximum and turbulence generation takes place mainly close to the bottom. %
Then, turbulence spreads towards the irrotational region but meanwhile it decays and there are phases of the cycle such that a laminar like flow is almost recovered. %
Even though the intensity of the dimensionless turbulent kinetic energy does not sensibly increase in $\rund$ with respect to that observed for $\runb$ (cf. figure~\ref{figY}), the larger value of the Reynolds number cause significant turbulent fluctuations to appear farther above the bed surface. %
%
%%%% Figure Y  *****************************************************
% ******************************************************************
%%%% Figure 3  *****************************************************
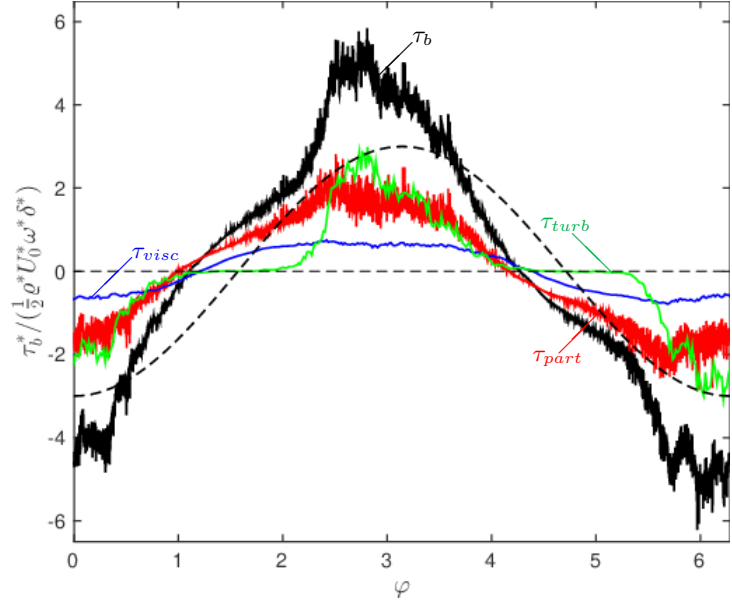
\begin{figure}[!]
\begin{picture}(0,210)(0,0)
  \ifthree
  \put(50,-4.5){\includegraphics[trim=0cm 1cm 0cm 0cm, clip, width=.7\textwidth]{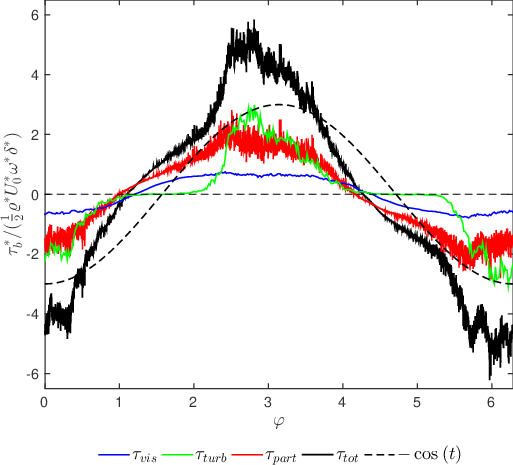}}
  \put(0,64){%
  \put(185,130){\line(1,1){15}}
  \put(200,145){\small$\tautot$}
  }%
  \put(0,29){%
  \put(275,93){\color{green!70!blue}\line(-1,1){15}}
  \put(248,110){\color{green!70!blue}\small$\taut$}
  }%
  \put(0,25){%
  \put(80,87.8){\color{blue!}\line(1,1){15}}
  \put(95,102.8){\color{blue!}\small$\tauv$}
  }%
  \put(0,22){%
  \put(255,70){\color{red!}\line(1,1){15}}
  \put(245,67){\color{red!}\small$\taup$}
  }%
  \fi
\end{picture}
\caption{%
The dimensionless value of $\tautot$ plotted versus the phase $\phase$ during the second cycle for $\Rdel=750$ and $\ds=%\frac{\ds^*}{\del^*}
0.335$ ($\runb$). %
The viscous contribution, the turbulent contribution and the contribution due to flow-particle interaction are also plotted along with the qualitative behaviour of the external velocity (broken line). %
}%
\label{fig3}
\end{figure}
%%
%%%% Figure 3  *****************************************************
\begin{figure}
\begin{picture}(0,210)(0,0)
  \ifthree
  \put(50,-4.5){\includegraphics[trim=0cm 1cm 0cm 0cm, clip, width=.7\textwidth]{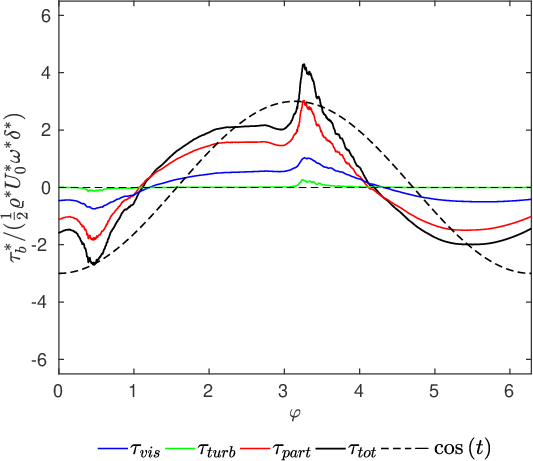}}
  \put(18,52){%
  \put(185,130){\line(1,1){15}}
  \put(200,145){\small$\tautot$}
  }%
  \put(-70,-8){%
  \put(266,114){\color{green!70!blue}\line(1,1){15}}
  \put(248,110){\color{green!70!blue}\small$\taut$}
  }%
  \put(0,25){%
  \put(80,87.8){\color{blue!}\line(1,1){15}}
  \put(95,102.8){\color{blue!}\small$\tauv$}
  }%
  \put(0,14){%
  \put(255,70){\color{red!}\line(1,1){15}}
  \put(245,67){\color{red!}\small$\taup$}
  }%
  \fi
\end{picture}
\caption{%
The dimensionless value of $\tautot$ and its contributions plotted versus the phase $\phase$ during the second cycle for $\Rdel=750$ and $\ds= 0.335$ ($\runbf$). %
}%
\label{fig3c}
\end{figure}
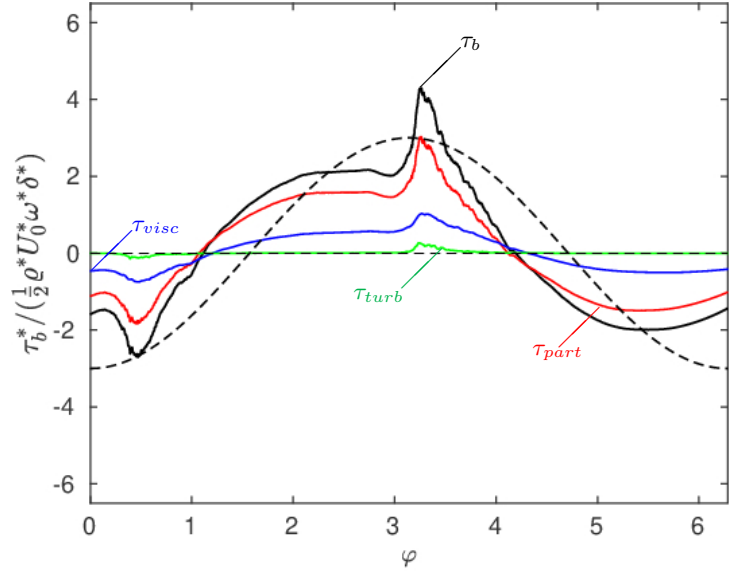
%%%% Figure 3a (figura definita 10 da Marco)  *****************************************************
\begin{figure}[!]
\begin{picture}(0,210)(0,0)
  \iften
  \put(80,0){\includegraphics[trim=0cm 0cm 0cm 0cm, clip, width=.6\textwidth]{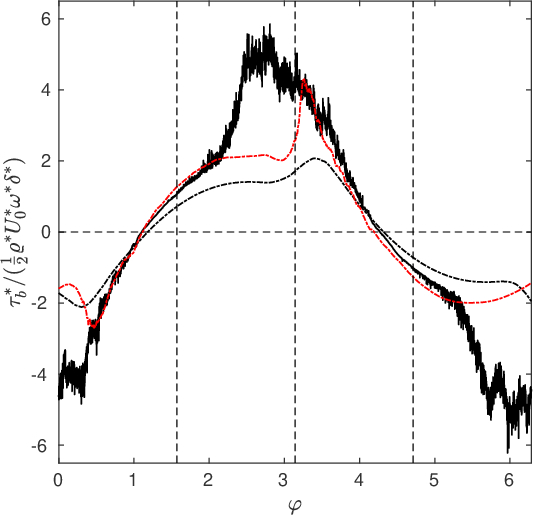}}
  \put(135,30){\footnotesize \rotatebox{90}{$\phase=\dfrac{1}{2}\pi$}}
  \put(195,30){\footnotesize \rotatebox{90}{$\phase=\pi$}}
  \put(238,30){\footnotesize \rotatebox{90}{$\phase=\dfrac{3}{2}\pi$}}
  \fi
\end{picture}
\caption{%
The dimensionless value of $\tautot$ plotted versus the phase $\phase$ during the second cycle of $\runb$ (mobile particle, solid line) and $\runbf$ (fixed particles, red dashdotted line), i.e. for the values of the parameters $\Rdel=750$ and $\ds=0.335$. %
Moreover, the black broken line indicates the values obtained in a run at $R_\delta=775$ over a smooth wall \citep[]['s results]{Mazzuoli2011}. %
}%
\label{fig3b}
\end{figure}
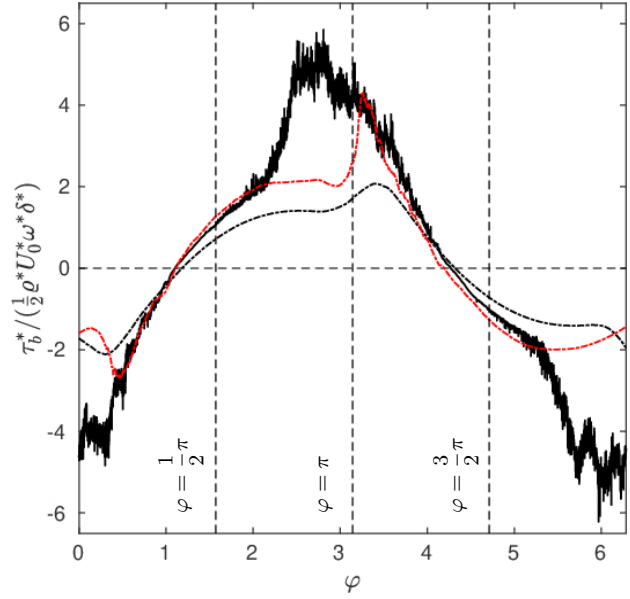
% ******************************************************************
%%%% Figure 3a (figura definita 10 da Marco)  *****************************************************
\begin{figure}[!]
\begin{picture}(0,120)(0,0)
  \iften
  \put(0,0){\includegraphics[trim=3cm 6cm 4cm 2cm, clip, width=.32\textwidth]{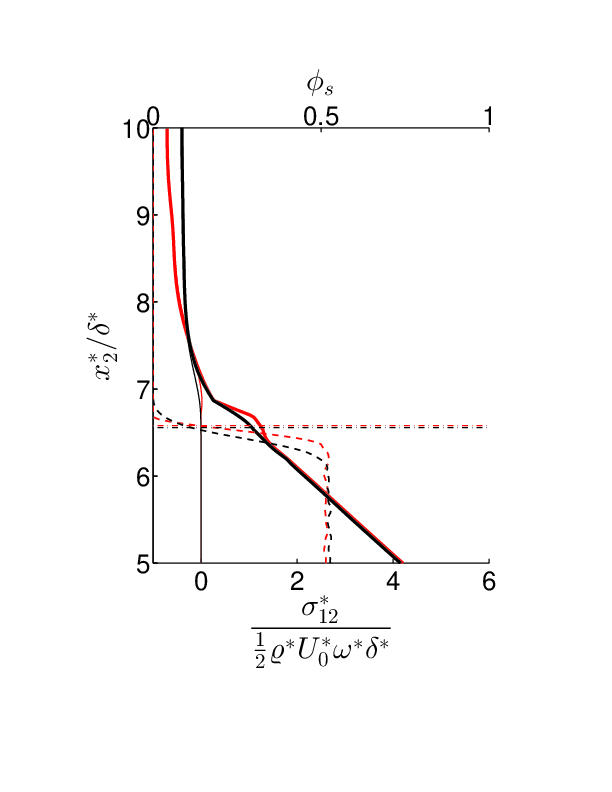}}
  \put(130,0){\includegraphics[trim=3cm 6cm 4cm 2cm, clip, width=.32\textwidth]{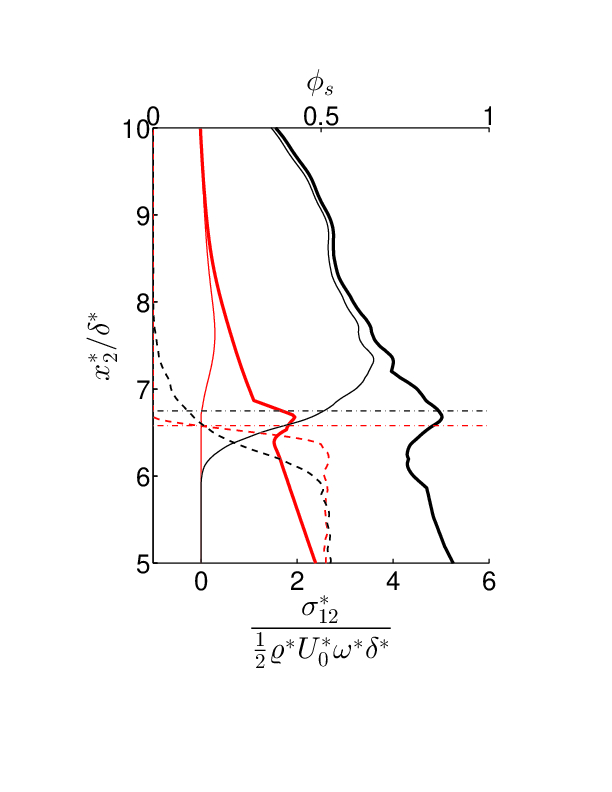}}
  \put(260,0){\includegraphics[trim=3cm 6cm 4cm 2cm, clip, width=.32\textwidth]{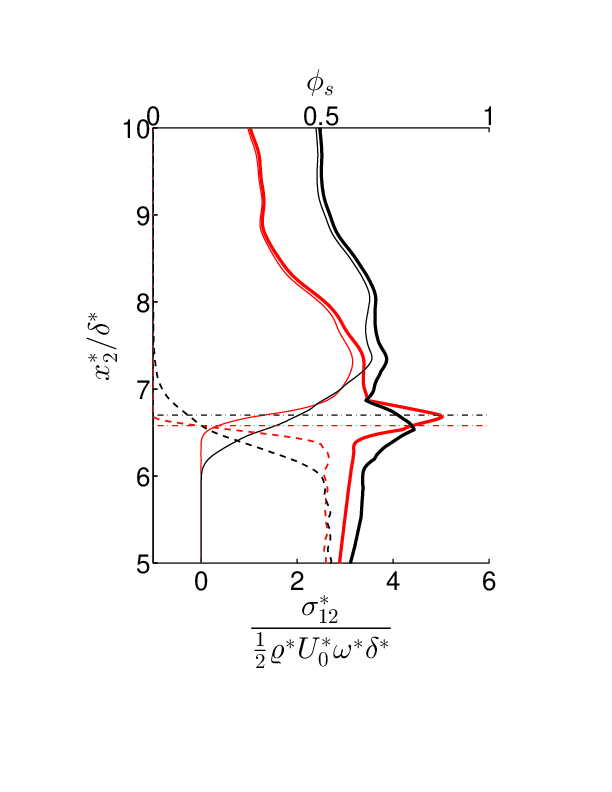}}
  \put(-5,145){$(a)$}
  \put(125,145){$(b)$}
  \put(255,145){$(c)$}
  \fi
\end{picture}
\caption{%
The dimensionless value of $\tautot$ (solid thick lines), of the Reynolds shear stress (solid thin lines) and the value of the solid volume fraction, $\svf$ (broken lines), plotted versus the wall-normal coordinate at the phases $(a):\:\phase=0.5\,\pi$, $(b):\:\phase=0.9\,\pi$ and $(c):\:\phase=1.035\,\pi$ of $\runb$ (mobile particles, black lines) and $\runbf$ (fixed particles, red lines), i.e. for the values of the parameters $\Rdel=750$ and $\ds=0.335$. %
Dashdotted horizontal lines indicate the bed surface elevation, $x_{2\,bottom}$, where $\svf=0.1$. %
}%
\label{fig3d}
\end{figure}
% ******************************************************************

\smallskip
\subsection{Evaluation of the bed shear stress}
Since the velocity increases rapidly above the bed surface and the presence of particles above the bed surface is limited to a thin layer (cf. figure~\ref{figX}), it is reasonable to suppose that the sediment flow rate is closely related to the bed shear stress. %
Figure~\ref{fig3} and \ref{fig3c} show the time development of the dimensionless streamwise component $\sigma_{12}=\frac{\sigma_{12}^*}{\frac{1}{2}\rho^*U^*_0 \delta^*\omega^*}$ of the averaged force per unit area exerted by the flow on the instantaneous bottom surface for $\runb$ and $\runbf$. %
The reader should notice that $\sigma_{12}^*(\xf{2}^*,t^*)$, evaluated at the bed surface ($\xf{2}^*=x_{2\,bottom}^*$) coincides with what is commonly defined as the bottom shear stress $\tautot^*$. %
Notwithstanding the fact that the force per unit area is averaged over the bottom surface, the value of $\tautot=\frac{\tau^*_b}{\frac{1}{2}\rho^*U^*_0 \delta^*\omega^*}$ is characterised by the presence of small random oscillations. %
To remove them, it would be necessary either to consider a much longer and wider computational box or to simulate a large number of cycles and to compute the phase averaged value. %
The oscillations of the force per unit area are defined ``small'' when compared with the oscillations which are observed when the value of $\tau_b$ is averaged over a much smaller horizontal surface. %
The value $\hat \tautot$ of $\sigma_{12}$ averaged over a portion of the instantaneous bottom surface which is $4~\del^*$ long, $2~\del^*$ wide and centered around the point $\left( x_1^*, x_3^*\right) = \left(\frac{\Lx}{2},\frac{\Lz}{2} \right) \del^*$ was computed to verify this point (the time development of $\hat \tautot$ is shown in a figure as\emph{supplementary material}). %

Three contributions to the value of $\tautot$ appearing in figures \ref{fig3} and \ref{fig3c} can be identified \citep{uhlmann2008,mazzuoli2018a}: %
$(i)$ the contribution due to the viscous stress ($\tau_{visc}$), %
$(ii)$ the contribution due to the turbulent stress ($\tau_{turb}$), %
$(iii)$ the contribution due to the flow-particle interactions ($\tau_{part}$). %
The procedure used to compute the different contributions is described in more details by \citet{mazzuoli2018a} and \citet{mazzuoli2019}. %
The time development of $\tautot$, for the $\runbf$ qualitatively agrees with that measured by \citet{jensen1989} (see figure \ref{fig1}) even though the Reynolds number of the laboratory experiment is somewhat different from that of the numerical simulation but, more importantly, the bottom of the experimental apparatus was smooth. %
Of course, the sediment motion and, in particular, the saltating grains greatly affect turbulence dynamics. %
Indeed, in $\runb$, the largest contributions to the bottom shear stress are those due to the turbulent stresses and the flow-particle interaction. %

To gain an idea of the role of the resting/moving sediment grains on turbulence dynamics, figure \ref{fig3b} shows the value of $\tautot$ computed for $\runb$ and $\runbf$, and that computed by \citet{Mazzuoli2011,mazzuoli2011a} who made a DNS of the OBL over a smooth wall for $R_\delta=775$. %
Even though the results of \citet{Mazzuoli2011,mazzuoli2011a} were obtained for a value of the Reynolds number slightly larger than that of the present simulations, figure \ref{fig3b} shows that the bottom shear stress of the smooth bottom case is significantly smaller than that of $\runbf$, where the particles were fixed and arranged in a plane-bed configuration. %
However, the phase when the inception of turbulence occurs and, in general, the time development of $\tautot$ are fairly comparable in these two cases. %
In particular, both in the simulation of \citet{Mazzuoli2011,mazzuoli2011a} and in $\runbf$, turbulence appears when the free stream velocity is decelerating. %
On the other hand, the presence of mobile sediments enhances the effect of the sediment on the transition process. %
This fact is clearly shown by figure~\ref{fig3d}, where the total dimensionless shear stress $\sigma_{12}$, the Reynolds shear stress, and the solid volume fraction for $\runb$ and $\runbf$ are plotted as functions of the wall-normal coordinate at flow reversal $(a)$ and at the phases when $\tautot$ is maximum in $\runb$ $(b)$ and in $\runbf$ $(c)$. %
The remarkable contribution of the Reynolds shear stress to $\tautot$, which is associated with the presence of moving particles, can be observed in figure~\ref{fig3d}b. %
Indeed, in semi-dilute and dense sheared suspensions (i.e. for $\svf>0.05$), the number of particles per unit volume that are exposed to the core flow is large enough to experience frequent contacts and generate significant overall drag. %
In order to understand the increase of the apparent roughness due to the particle motion, it is worth to note that in a suspension of mono-sized spheres at $\svf=0.1$, the distance between particles is approximately equal to their diameter and the flow resistance in steady condition is a few times larger than that attained with the tightest arrangement of the spheres, i.e. in the ``cannonball'' configuration. %
For instance, \citet{schlichting1936}, who carried out experiments on channel flow over a plane layer of spheres of diameter $\ds^*$ arranged at the vertices of hexagons of side $\ell^*$, found that the equivalent roughness obtained for $\ell^*$ equal to $2~\ds^*$ was approximately five times larger than that for closely packed spheres. %

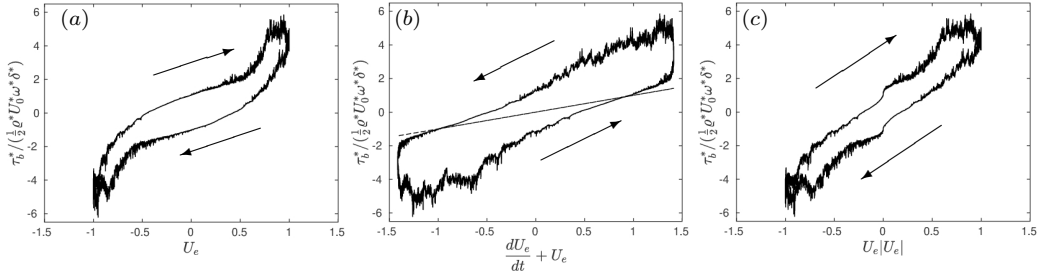
\begin{figure}[!]
\begin{picture}(0,105)(0,0)
  \iffour
  \put(-5,1.5){\includegraphics[trim=0cm 0cm 0cm 0cm, clip, width=.33\textwidth]{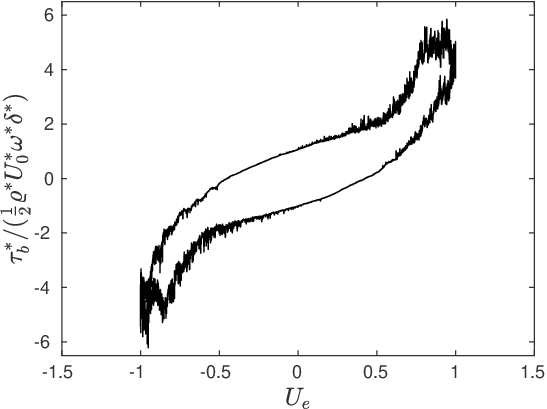}}
  \put(124,-4.5){\includegraphics[trim=0cm 0cm 0cm 0cm, clip, width=.33\textwidth]{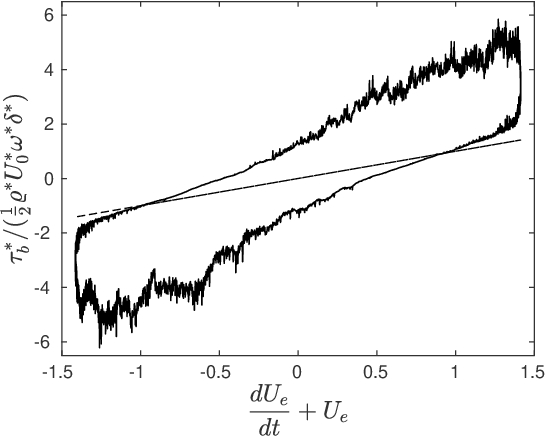}}
  \put(254,1){\includegraphics[trim=0cm 0cm 0cm 0cm, clip, width=.33\textwidth]{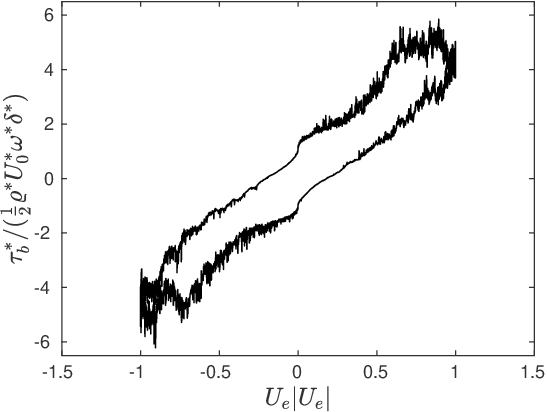}}
  \put(15,88){\footnotesize $(a)$}
  \put(140,88){\footnotesize $(b)$}
  \put(270,88){\footnotesize $(c)$}
  \put(90,49){\vector(-3,-1){30}}
  \put(50,69){\vector(3,1){30}}
  \put(195,37){\vector(2,1){30}}
  \put(200,82){\vector(-2,-1){30}}
  \put(345,50){\vector(-3,-2){30}}
  \put(298,64){\vector(3,2){30}}
  \fi
\end{picture}
\caption{%
The dimensionless value of $\tautot$ plotted both versus the free stream velocity $\Ue$ (panel $a$), the quantity $\left( \frac{d\Ue}{dt}+\Ue \right)$ (panel $b$) and the quantity $\Ue\vert\Ue\vert$ (panel $c$) for $\Rdel=750$ and $\ds=0.335$ ($\runb$). %
In the laminar regime, the curve $\tautot$ versus $\left( \frac{d\Ue}{dt}+\Ue \right)$ is the straight line crossing the origin of the axes and plotted in panel $b$. %
}%
\label{fig4}
\end{figure}

Because of turbulence growth, which takes place when the external flow is close to its largest values, and its subsequent damping, the values of $\tautot$ attained during flow acceleration differ from those observed during flow deceleration, even though the free stream velocity is equal. %
Hence, if $\tautot$ is plotted versus the free stream velocity $\Ue$, a hysteresis orbit can be observed (see figure \ref{fig4}a). %
A similar orbit is observed even considering the curve $\tautot$ versus the dimensionless quantity $\left( \Ue + \frac{d\Ue}{dt} \right) $, which in the laminar case should be a straight line passing through the origin of the axes, and the dimensionless quantity $\Ue\vert\Ue\vert$. %
These results show that, even though empirical relationships do exist which allow a reliable estimate of the friction factor $\frifac=\taumax^*/\left(\frac{1}{2}\densf^*\U^{*2}\right)$ and it is relatively easy to predict the maximum value $\tau^*_{b,max}$ of the bottom shear stress, the time development of bottom shear stress during the flow cycle is much more difficult to be predicted. %
In order to fulfil the objectives (\emph{I}) and (\emph{II}) stated in \S~$1$, let us see in the following section how the bed shear stress is related to the sediment flow rate. %
%
 
%%%% Figure 4  *****************************************************
% ******************************************************************
%%%% Figure 5  *****************************************************
\begin{figure}
\begin{picture}(0,170)(0,0)
  \iffive
\put(0,0){%
  \put(0,0){\includegraphics[trim=0cm 0cm 0cm 0cm, clip, width=.49\textwidth]{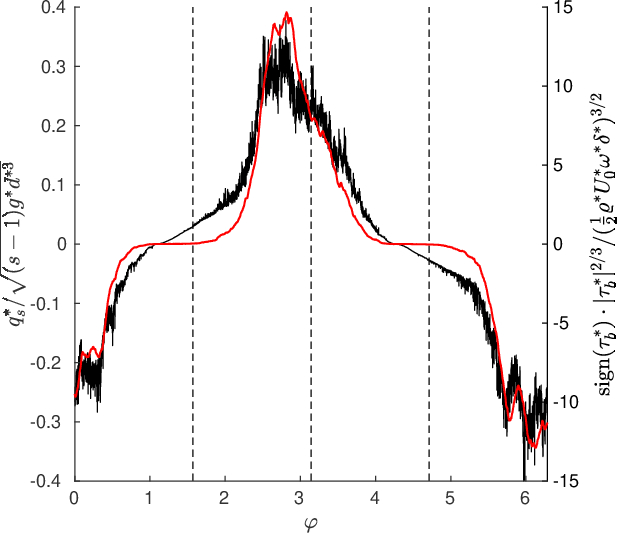}}
  \put(38,30){\footnotesize \rotatebox{90}{$\phase=\dfrac{1}{2}\pi$}}
  \put(85,30){\footnotesize \rotatebox{90}{$\phase=\pi$}}
  \put(110,30){\footnotesize \rotatebox{90}{$\phase=\dfrac{3}{2}\pi$}}
  \put(-5,155){$(a)$}
}%
\put(200,0){%
  \put(0,0){\includegraphics[trim=0cm 0cm 0cm 0cm, clip, width=.48\textwidth]{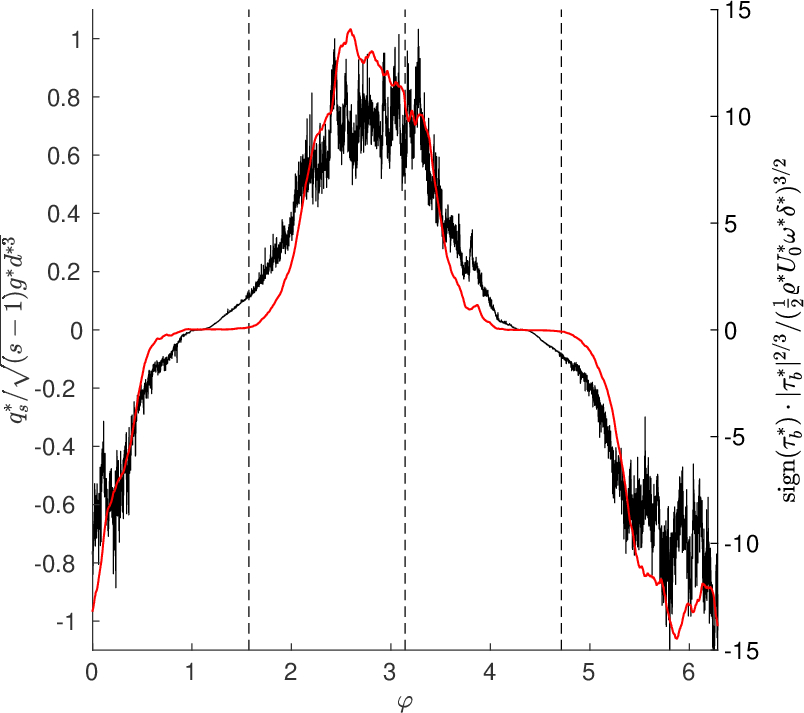}}
  \put(38,30){\footnotesize \rotatebox{90}{$\phase=\dfrac{1}{2}\pi$}}
  \put(85,30){\footnotesize \rotatebox{90}{$\phase=\pi$}}
  \put(110,30){\footnotesize \rotatebox{90}{$\phase=\dfrac{3}{2}\pi$}}
  \put(-5,155){$(b)$}
}%
  \fi
\end{picture}
\caption{%
Values of $\frac{\qpmean^*}{\sqrt{\left(\s-1\right) \g^* \ds^{*3} }}$ (thick red line) and the signed values of $\left(\frac{\vert\tautot^*\vert}{(\frac{1}{2}\densf^*\U^*\om^*\del^*}\right)^{3/2}$ (thin black line) plotted versus the phase $\phase$ during the second cycle for $\ds=0.335$, $(a)$ $\Rdel=750$ ($\runb$) and $(b)$ $\Rdel=1000$ ($\rund$). %
}%
\label{fig5}
\end{figure}
% ******************************************************************
%%%% Figure 6  *****************************************************
\begin{figure}
\begin{picture}(0,200)(0,0)
  \ifeight
  \put(50,-85){\includegraphics[trim=0cm 0cm 0cm 0cm, clip, width=.75\textwidth]{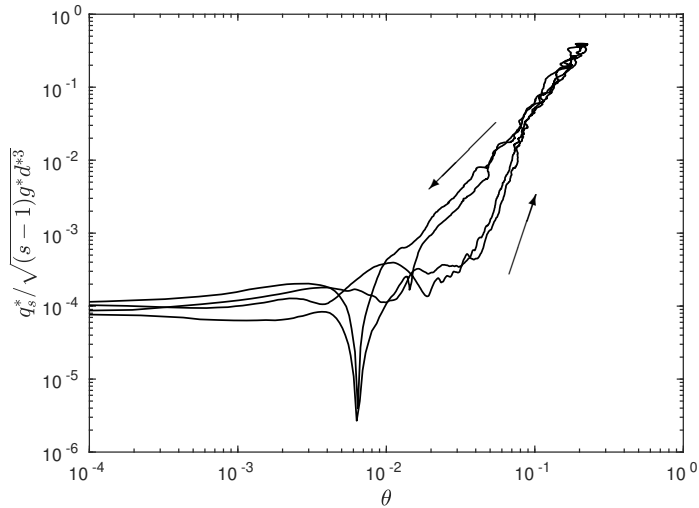}}
  \put(240,88){\vector(1,3){10}}
  \put(235,145){\vector(-1,-1){25}}
  \fi
\end{picture}
\caption{%
Dimensionless sediment flow rate $\qpmean$ as a function of the Shields parameter $\shields$ for $\Rdel=750$ and $\ds=0.335$ ($\runb$). %
Arrows indicate the orbital direction. %
}%
\label{fig6}
\end{figure}

\smallskip
\subsection{Evolution of the sediment flow rate and dependence on the flow properties}
Figure~\ref{fig5} shows the dimensionless sediment transport rate per unit width $\qpmean=\frac{q_s^*}{\sqrt{\left(s-1\right)\g^*\ds^{*3}}}$ (i.e. the volume of sediment grains which crosses a plane $\xf{1}=\mbox{constant}$ per unit time and width), averaged over the ``homogeneous'' directions $\xf{1}$ and $\xf{3}$, as a function of the phase $\phase$ during the second cycle. %
Since it is a common practice to correlate the sediment flow rate to the power $3/2$ of the bottom shear stress, in the same figure the signed value of $\vert\tautot\vert^{3/2}$ is also plotted. %
Figure~\ref{fig5} clearly shows that a fair correlation does exist between $\qpmean$ and $\vert\tautot\vert^{3/2}$ for both $\runb$ (panel $(a)$) and $\rund$ (panel $(b)$), even though analogous differences can be observed in the two runs. %
Taking into account that $(i)$ turbulence intensity during the decelerating phases of the cycle is different from that observed during the accelerating phases and $(ii)$ the sediment particles are set into motion and transported more easily when turbulence intensity is high, it is reasonable to expect values of $\qpmean$ during the late accelerating phases larger than those observed during the decelerating phases, even for the same value of the bottom shear stress. %
In figure~\ref{fig6}, the value of $\qpmean$ is plotted versus the Shields parameter $\shields = \frac{\tautot^*}{\left(\denss^*-\densf^*\right) \g^* \ds^*}$, describing two nearly coincident orbits during one oscillation period. %
In fact, for values of the Shields parameter slightly larger than its critical value ($10^{-2}\lesssim\shields\lesssim 10^{-1}$), the sediment transport rate depends not only on $\shields$ but also on the value of $d\shields/dt$. %
A similar finding was obtained by \citet{Vittori2003} who showed that, in an oscillatory flow, the amount of sediment grains picked up from the bed and carried into suspension correlates better with the production of turbulent kinetic energy than with the Shields parameter. %
The results plotted in figure~\ref{fig6} show that the sediment transport rate tends to vanish for a finite value of $\shields$ while it assumes a small but finite value for $\shields$ tending to zero. %
To understand this behaviour of $\qpmean$, the reader should take into account that even for vanishing values of $\shields$ the sediments keep moving mainly because moving particles take some time to find a stable position on the bed surface \citep{clark2017} and also because of the effects of the imposed pressure gradient \citep{mazzuoli2019}. %
Hence, the sediment transport tends to vanish for small values of $\shields$ when the effects of the pressure gradient and the viscous forces on the particles balance inertial effects. %
On the other hand, the results plotted in figure \ref{fig6} show that fair predictions of the sediment transport rate can be obtained by assuming that $q_s$ depends only on $\theta$ when the Shields parameter assumes relatively large values. %

Some further insight into the effects of turbulence dynamics on sediment motion can be obtained looking at the time development of the vortex structures and at the associated dynamics of sediment grains. %
%
%%%% Figure 13  *****************************************************
\begin{figure}
\begin{picture}(0,400)(0,5)
  \ifthirteen
    \put(-5,300){\includegraphics[trim=1cm 2cm 0cm 2cm, clip, width=.52\textwidth]{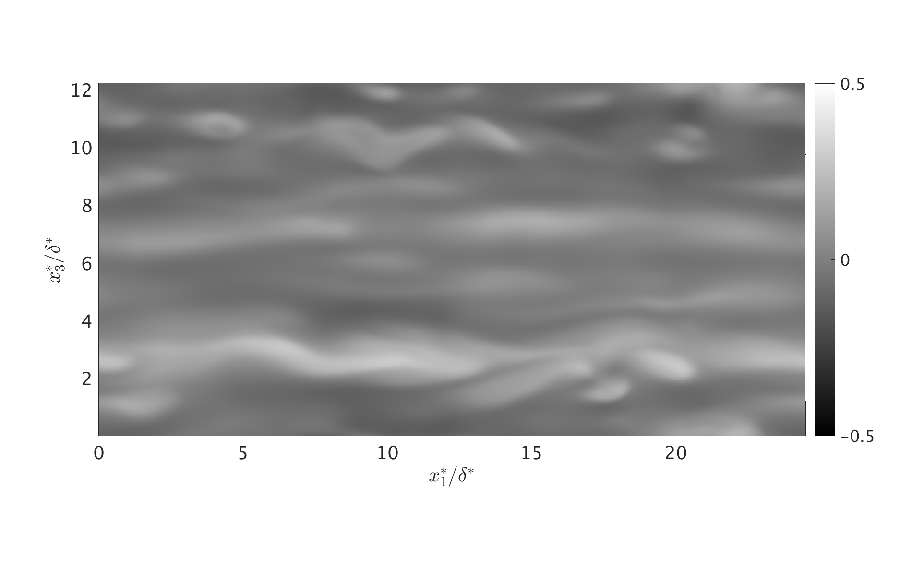}}
    \put(-8,390){\small$(a)$}
    \put(190,300){\includegraphics[trim=1cm 2cm 0cm 2cm, clip, width=.52\textwidth]{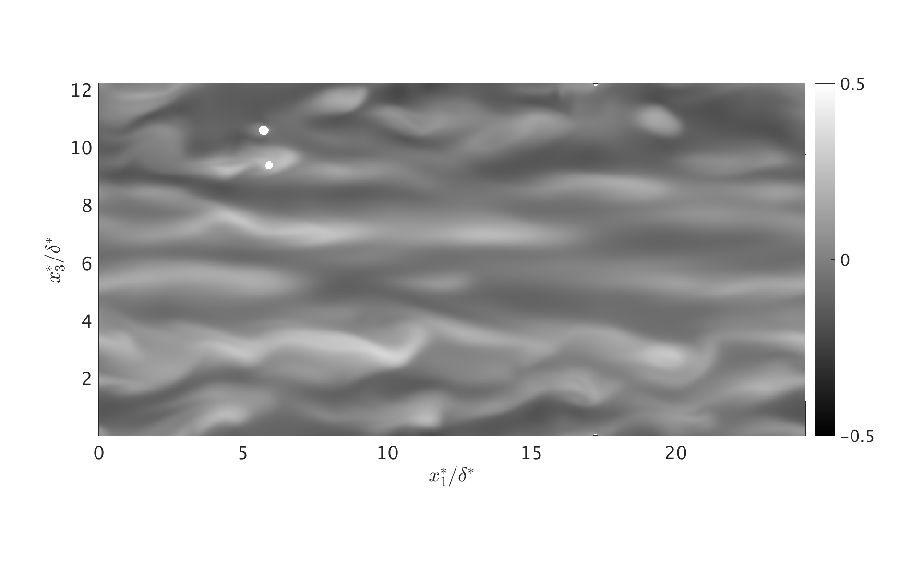}}
    \put(188,390){\small$(b)$}
    \put(-5,200){\includegraphics[trim=1cm 2cm 0cm 2cm, clip, width=.52\textwidth]{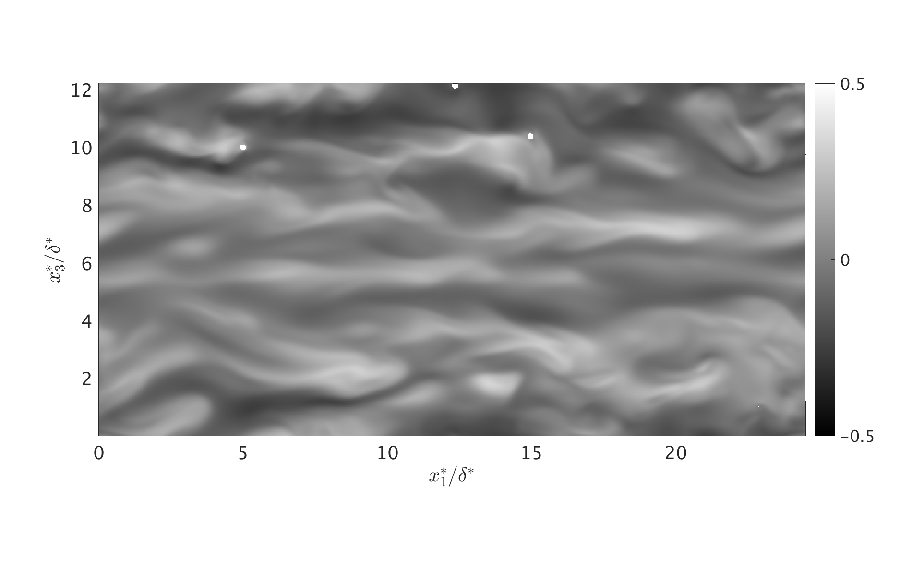}}
    \put(-8,290){\small$(c)$}
    \put(190,200){\includegraphics[trim=1cm 2cm 0cm 2cm, clip, width=.52\textwidth]{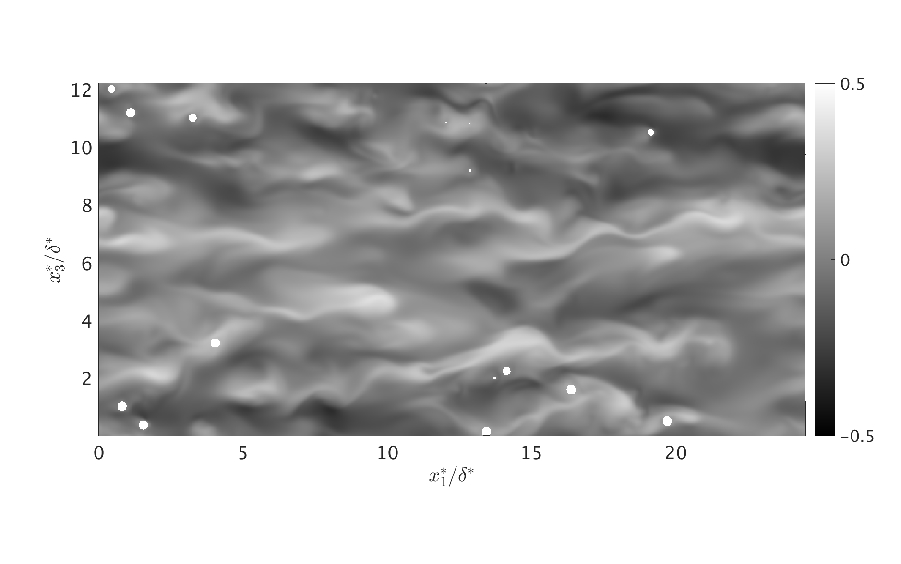}}
    \put(188,290){\small$(d)$}
    \put(-5,100){\includegraphics[trim=1cm 2cm 0cm 2cm, clip, width=.52\textwidth]{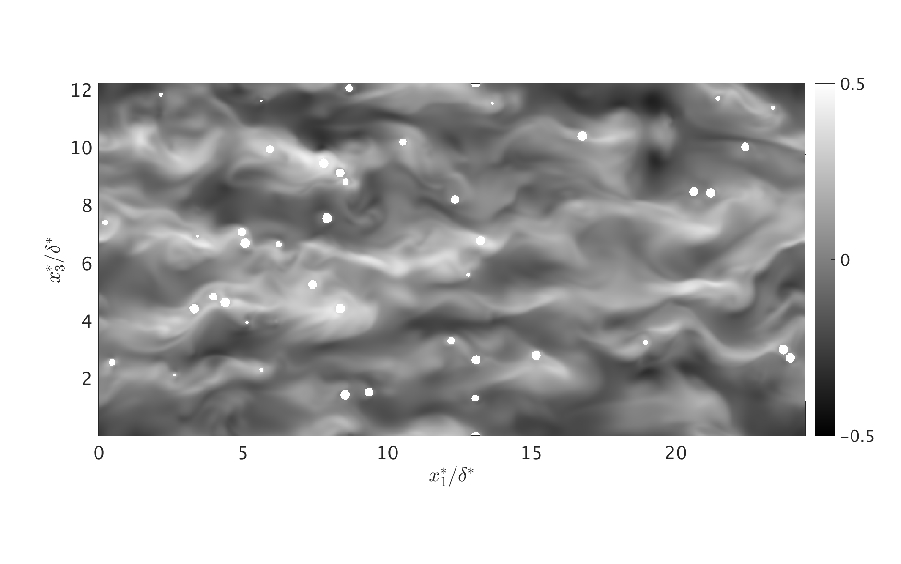}}
    \put(-8,190){\small$(e)$}
    \put(190,100){\includegraphics[trim=1cm 2cm 0cm 2cm, clip, width=.52\textwidth]{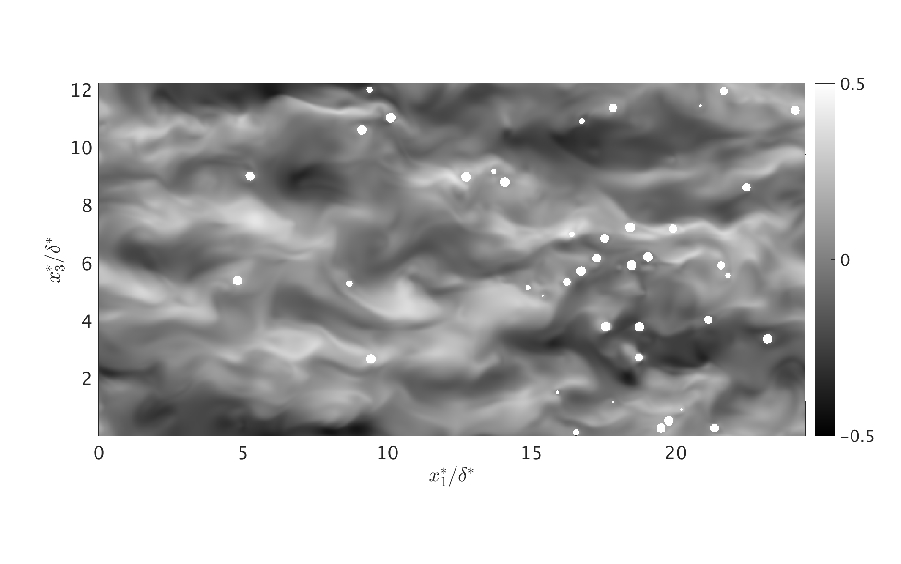}}
    \put(188,190){\small$(f)$}
    \put(-5,0){\includegraphics[trim=1cm 2cm 0cm 2cm, clip, width=.52\textwidth]{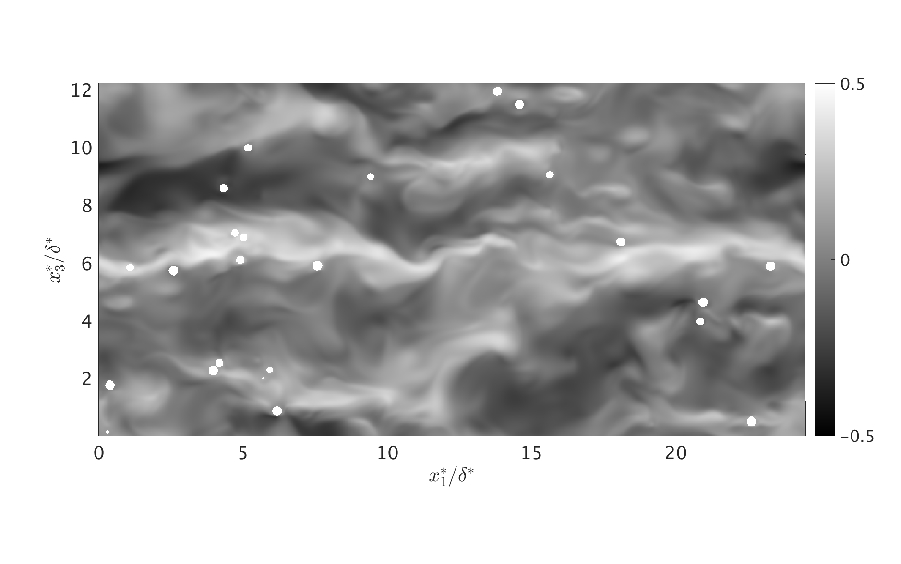}}
    \put(-8,90){\small$(g)$}
    \put(190,0){\includegraphics[trim=1cm 2cm 0cm 2cm, clip, width=.52\textwidth]{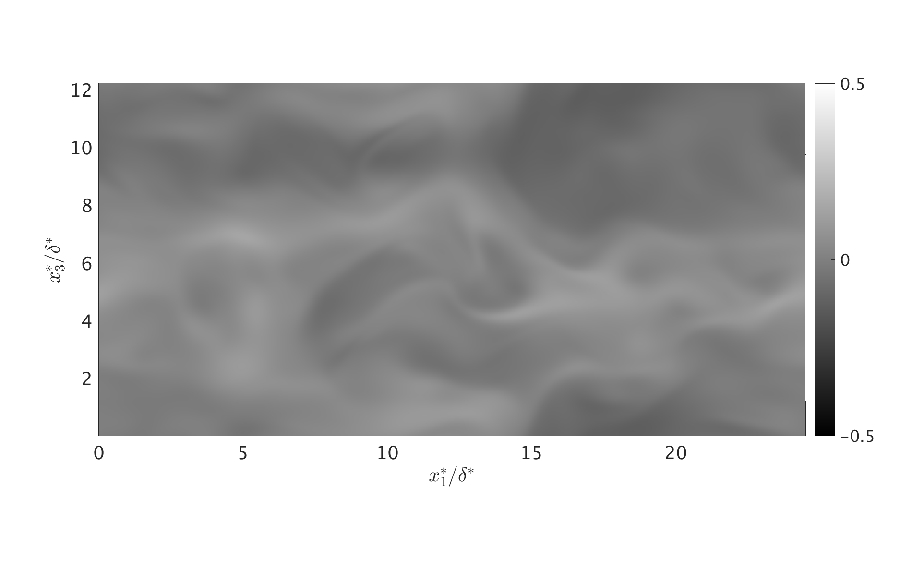}}
    \put(188,90){\small$(h)$}
  \fi
\end{picture}
\caption{%
Streamwise velocity fluctuations at plane $\xf{2}=7.5$ ($\xf{2}-x_{2,bottom}=0.9$) at different phases of the cycle: (a) $\om^*t^*=1.73\pi$ ($u_\tau^*\ds^*/\nu^*=15$, $\shields=0.12$), (b) $\om^*t^*=1.75\pi$ ($u_\tau^*\ds^*/\nu^*=16$, $\shields=0.13$), (c) $\om^*t^*=1.78\pi$ ($u_\tau^*\ds^*/\nu^*=18$, $\shields=0.16$), (d) $\om^*t^*=1.80\pi$ ($u_\tau^*\ds^*/\nu^*=19$, $\shields=0.18$), (e) $\om^*t^*=1.85\pi$ ($u_\tau^*\ds^*/\nu^*=22$, $\shields=0.24$), (f) $\om^*t^*=1.91\pi$ ($u_\tau^*\ds^*/\nu^*=20$, $\shields=0.20$), (g) $\om^*t^*=2.05\pi$ ($u_\tau^*\ds^*/\nu^*=19$, $\shields=0.17$), (h) $\om^*t^*=2.40\pi$ ($u_\tau^*\ds^*/\nu^*=6$, $\shields=0.02$). %
$\Rdel=750, \ds=0.335$ ($\runb$). %
The full sequences of visualisations of streamwise velocity and spanwise vorticity fluctuations at plane $\xf{2}=7.5$ can be found online in the movie no. 1 and movie no. 2 of the \textit{supplementary material}, respectively. %
}%
\label{fig13}
\end{figure}
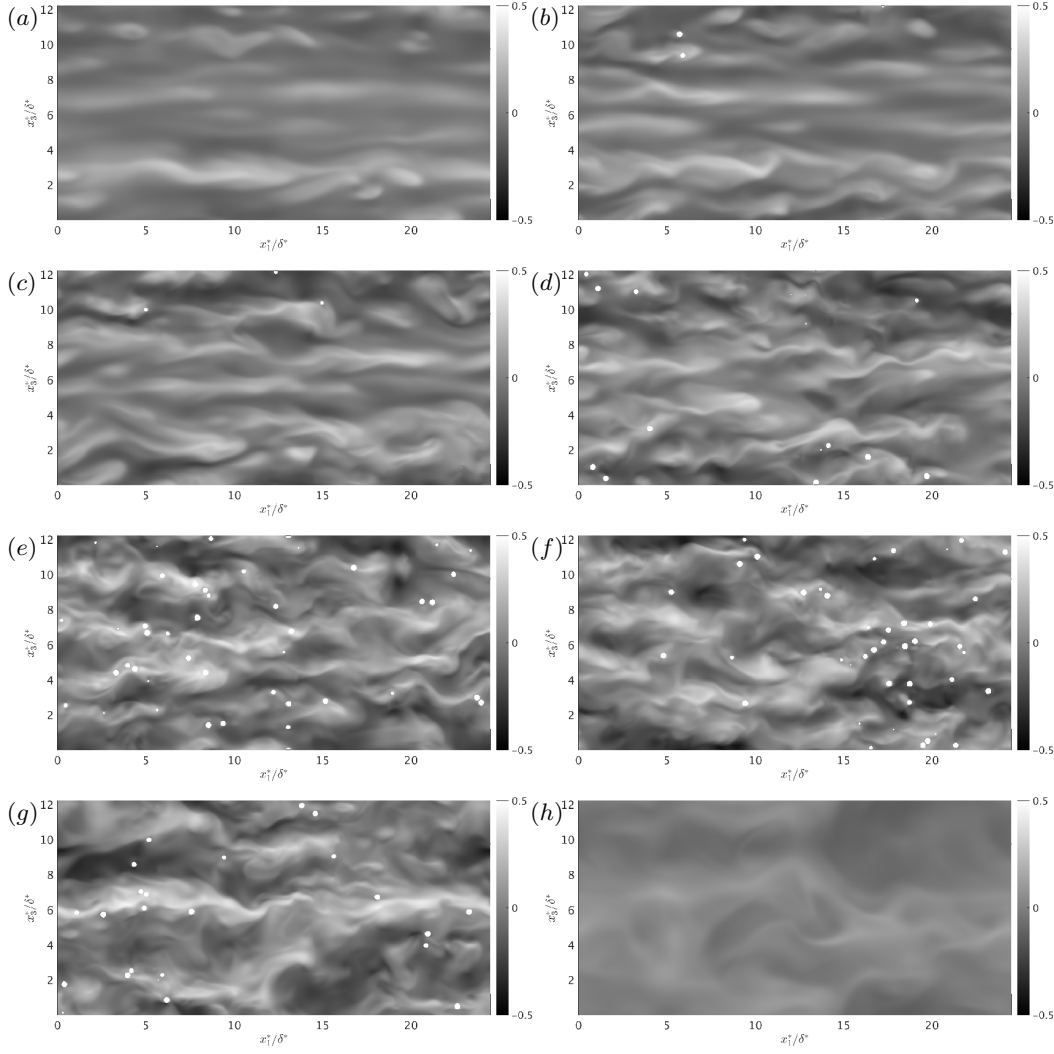
% ******************************************************************
%
Figure~\ref{fig13} shows the streamwise velocity fluctuations in the horizontal plane $\xf{2}-x_{2\,bottom}=0.9$, along with sediment particles (white dots) that are picked up from the bottom and, during their motion, cross this plane. %
The panel $(a)$ of the figure shows that low and high speed streaks characterise the flow field during the flow acceleration. %
Later, as also found by \citet{Costamagna2003,Mazzuoli2011,mazzuoli2016b,mazzuoli2019b} who simulated an oscillatory boundary layer over a smooth bottom, the streaks oscillate, twist and interact (see panels $(b)$ and $(c)$ of figure~\ref{fig13}) and, then, they break generating small vortex structures and a fully turbulent flow (see panels $(d)$, $(e)$, $(f)$ and $(g)$). %
Eventually, the turbulent eddies decay because of viscous effects and the flow recovers a laminar like behaviour (see panel $(h)$ of figure~\ref{fig13}). %
When the external velocity is maximum, also the turbulence level is high and a large amount of sediment particles are picked up from the bottom and transported by the flow. %
Then, at the late stages of flow deceleration, the turbulent eddies damp out, the bottom shear stress tends to vanish and the sediment particles settle down. %
The interested reader can look at the movie which is available as \textit{supplementary material}. %

We have discussed the flow field and sediment dynamics during the second cycle since, for $\Rdel=750$, the average results obtained during this cycle are similar to those obtained during the first cycle thus suggesting that the flow has attained its periodic status. %
The results of the numerical simulation carried out for $\Rdel=450$ and $\ds=0.335$ ($\runa$) allow to appreciate more easily the dynamics of both the vortex structures and the sediment particles because the process which leads to turbulence appearance is slower and similar to that observed experimentally over a smooth wall by \citet{Carstensen2010} who made flow visualizations for values of the Reynolds number close to its critical value. %
Indeed, Kajiura's (1968) criterion suggests that, for $\ds=0.335$, the critical value of the Reynolds number falls around $310$. %
Hence, for $\Rdel=450$, turbulence is expected  to be weak and large fluctuations of turbulence intensity from cycle to cycle are expected to be present. %
%

%
%%%%% Figure 11  *****************************************************
\begin{figure}
\begin{picture}(0,240)(0,0)
  \ifeleven
    \put(0,160){\includegraphics[trim=0cm .3cm 0cm -.1cm, clip, width=1\textwidth]{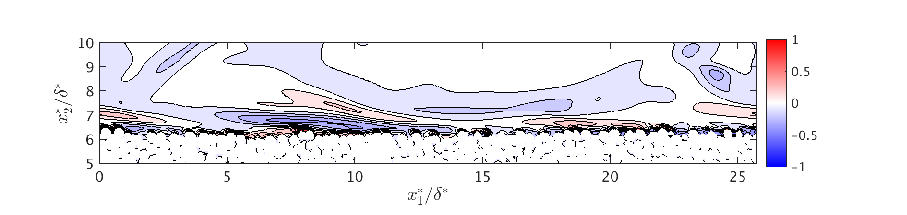}}
    \put(0,80){\includegraphics[trim=0cm .1cm 0cm .7cm, clip, width=1\textwidth]{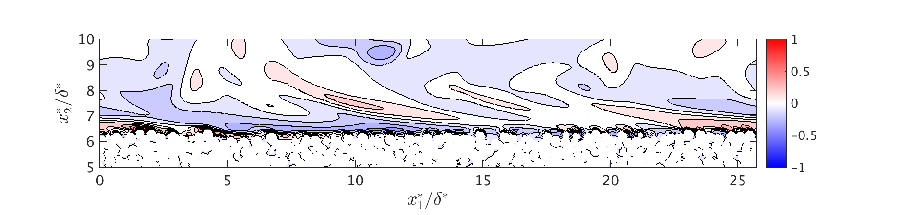}}
    \put(0,0){\includegraphics[trim=0cm .1cm 0cm .7cm, clip, width=1\textwidth]{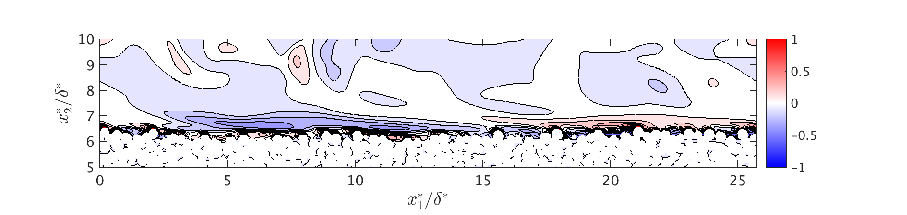}}
    \put(10,230){$(a)$}
    \put(10,150){$(b)$}
    \put(10,70){$(c)$}
  \fi
\end{picture}
\caption{%
Dimensionless fluctuating spanwise component of vorticity $\vort{3}^{'*}\del^*/\U^* = \vort{3}' = \vort{3}-\zxav{\vort{3}}$ at $\om^*t^*=5.7\pi$ and $(a)$ $\xf{3}=2$, $(b)$ $\xf{3}=6$, $(c)$ $\xf{3}=8$ for $\Rdel=450$ and $\ds=0.335$ ($\runa$). %
}%
\label{fig11}
\end{figure}
% ******************************************************************
Figure \ref{fig11} shows the spanwise vorticity component at three streamwise-vertical planes defined by $\xf{3}=2$, $\xf{3}=6$ and $\xf{3}=8$ and at $t=5.7\pi$, once the plane-averaged value is removed. %
The plots show that, close to the bottom, positive and negative regions alternate in the streamwise direction with a wavelength of approximately $12.5 \del^*$. %
%
%%%% Figure 12  *****************************************************
\begin{figure}
\begin{picture}(0,270)(0,0)
  \iftwelve
  \put(0,135){%
  \put(0,0){\includegraphics[trim=0cm 16.6cm 0cm 9.5cm, clip, width=.47\textwidth]{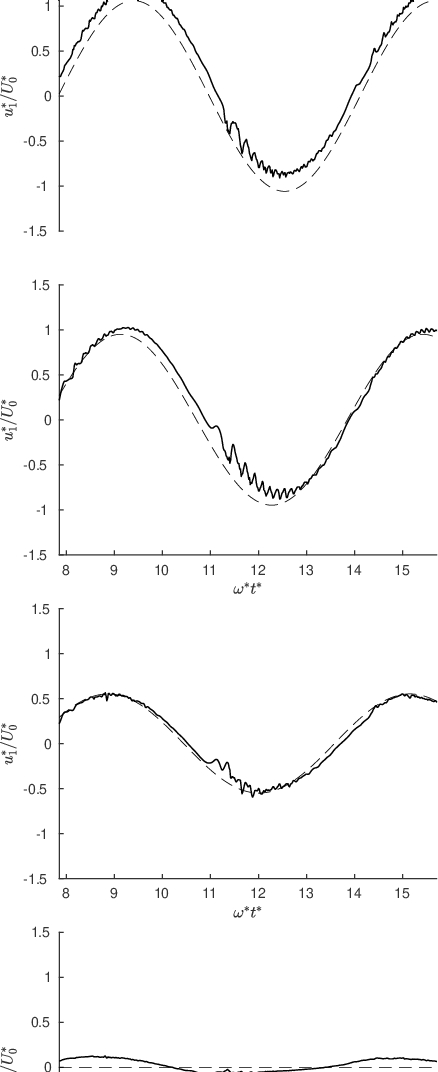}}
  \put(200,0){\includegraphics[trim=0cm 5.2cm 0cm 21cm, clip, width=.47\textwidth]{\figdir figure15ab}}
  \put(32,120){\footnotesize $(a)$}
  \put(232,120){\footnotesize $(b)$}
  }%
  \put(0,0){%
  \put(0,0){\includegraphics[trim=0cm 16.6cm 0cm 9.5cm, clip, width=.47\textwidth]{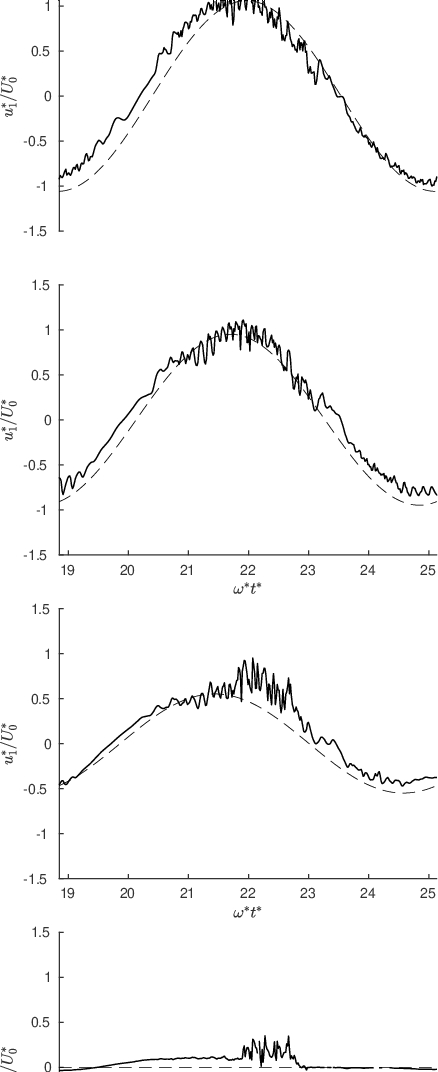}}
  \put(200,0){\includegraphics[trim=0cm 5.2cm 0cm 21cm, clip, width=.47\textwidth]{\figdir figure15cd}}
  \put(32,120){\footnotesize $(c)$}
  \put(232,120){\footnotesize $(d)$}
  }%
  \fi
\end{picture}
\caption{%
Streamwise velocity component plotted versus time for $\xf{1}=\Lx/2$, $\xf{3}=\Lz/2$ and $(a)$ $\xf{2}=x_{2,bottom}+1$, $(b)$ $\xf{2}=x_{2,bottom}+0.5$. %
Continuous line = numerical results; broken line = Stokes' solution. %
$\Rdel=450$ and $\ds=0.335$ ($\runa$). %
}%
\label{fig12}
\end{figure}
% ******************************************************************
These coherent spanwise vortex structures appear when the free stream velocity is almost maximum and, then, they are convected in the streamwise direction generating almost regular oscillations of the velocity field as it appears in figure~\ref{fig12}a,b, where the streamwise velocity component is plotted versus $\om^*t^*$ at two of the locations considered in figure \ref{fig2}. %
Later on, these vortex structures attain their maximum intensity and eventually decay because of viscous effects and disappear. %

As already pointed out, similar spanwise vortices were visualised by \citet{Carstensen2010} during laboratory experiments and by \citet{Mazzuoli2011,mazzuoli2011a}, who reproduced the experiments by \citet{Carstensen2010} by means of DNSs and confirmed that the first vortex structures which appear during the transition process are two-dimensional spanwise vortices. %
In fact, during the first oscillation cycles, notwithstanding the presence of the sediment particles which make the bottom rough and certainly affect the transition process (see i.a. \citet{blondeaux1991}), the present numerical findings show vortex structures which are qualitatively in agreement with both the predictions of the linear stability analysis of \citet{blondeaux1979} and the results of the DNSs of \citet{Vittori1998}, \citet{Costamagna2003} and \citet{bettencourt2018}, who considered a smooth bottom. %
\citet{blondeaux1979}, by using a momentary criterion of instability, showed that the laminar Stokes boundary layer is linearly and momentary unstable when the Reynolds number $\Rdel$ is larger than $86$ and the fastest growing mode is two-dimensional and characterised by a streamwise wavelength of about $12.5~\del^*$. %
For values of $\Rdel$ close to its critical value, the instability predicted by the linear analysis is restricted to a small part of the cycle and during the remaining parts of the cycle, the amplification rate becomes negative and the flow recovers a laminar like behaviour. %
Larger values of $\Rdel$ widen the unstable parts of the cycle and lead to larger perturbations. %
However, the flow is stable ``on the average''. %
Indeed the results of \citet{blennerhassett2002} show that small perturbations of the Stokes' solution are characterised by an averaged growth only when the Reynolds number is larger than $1416$. %
%

%%%% Figure 19a  *****************************************************
\begin{figure}
\begin{picture}(0,180)(0,0)
  \ifnineteen
  \put(50,0){\includegraphics[trim=0cm 0cm 0cm 0cm, clip, width=.7\textwidth]{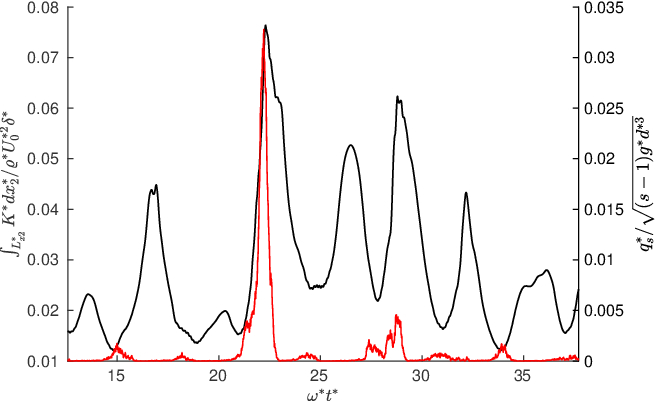}}
  \fi
\end{picture}
\caption{%
Time development of the dimensionless sediment flow rate $\qpmean$ (red line) and the turbulent kinetic energy $K$ (black line), integrated along the $\xf{2}$-direction from the bed surface to the top of the computational domain. %
$\Rdel=450$ and $d=0.335$ ($\runa$). %
}%
\label{fig19a}
\end{figure}
%% ******************************************************************
%%%% Figure 19  *****************************************************
\begin{figure}
\begin{picture}(0,190)(0,0)
  \ifnineteen
  \put(50,0){\includegraphics[trim=0cm 0cm 0cm 0cm, clip, width=.7\textwidth]{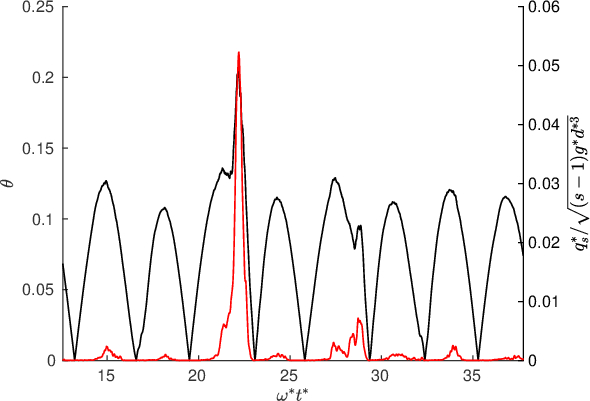}}
  \fi
\end{picture}
\caption{%
Time development of the dimensionless sediment flow rate $\qpmean$ (red line) and the Shields parameter $\shields$ (black line). %
$\Rdel=450$ and $d=0.335$ ($\runa$). %
}%
\label{fig19}
\end{figure}
%*****************************************************

Transition to turbulence is triggered by nonlinear effects, which can no longer be neglected when the perturbations attain large values during the momentary unstable phases, and it is also affected by wall imperfections \citep{blondeaux1994} and three-dimensional effects. %
Moreover, in the present simulations the resting/moving sediment grains certainly affect the transition process. %
Figure~\ref{fig12}c,d, where the streamwise velocity component is plotted versus $\om^*t^*$ for $\Rdel=450$ and $\ds=0.335$, shows that turbulent velocity fluctuations appear during the fourth cycle, being triggered by the vortex structures generated during the previous cycles, which decayed but were still strong enough to trigger the growth of large perturbations of the laminar flow. %
This peculiar behaviour of turbulence dynamics clearly appears in figure~\ref{fig19a} where the value of $K^*$, i.e. the plane-averaged turbulent kinetic energy per unit area of the bottom integrated over the whole computational domain, is plotted versus time. Indeed, large variations of $K^*$ from half-cycle to half-cycle can be observed in figure~\ref{fig19a}. %
During the phases characterized by the growth of $\int_{L^*_{x2}}K^* dx^*_2/\varrho^*U^{*2}_0\delta^*$, the sediment flow rate tends to increase but then it decreases even if the integral of $K^*$ still keeps growing. %
This results can be understood by observing that large values of $K^*$ far from the bottom are encountered later than close to the bottom, where turbulence is produced, as an effect of  the diffusion of turbulent fluctuations (see for example figure~\ref{figY} referred to $\runb$). %
Therefore, the maxima of the integral of $K^*$ lag behind the maxima of $K^*$ close to the bed, which are associated with large values of the bed shear stress and of the sediment flow rate. %
The fact that the sporadic inception to turbulence in $\runa$ is due to the presence of particles appears reasonable since \citet{Carstensen2010}, in absence of particles, could not observe the transition to turbulence for $\Rdel=450$. %
However, it is evident from the present results that the values of the parameters characterising $\runa$ lay on the edge of the intermittently-turbulent region of the parameter space. %

The value of the plane-averaged sediment flow rate $\qpmean$ per unit width is plotted in figure \ref{fig19} along with the value of the Shields parameter as function of time. %
First of all, it is worth pointing out that large fluctuations of the Shields parameter $\shields$ from cycle to cycle are present, because of the large fluctuations of turbulence intensity: the appearance of strong turbulent eddies close to the bottom gives rise to large velocity gradients at the bottom and large values of $\tau_b$. %
Moreover, as already pointed out, the values of $\qpmean$ during flow acceleration differ from the values during the flow deceleration, even though the Shields parameter assumes the same value. %
Hence, the results show that, for $\Rdel$ close to its critical value, a sediment transport rate predictor based on the assumption that $\qpmean$ depends only on $\shields$ cannot provide good predictions. %
If turbulence is strong, a large number of sediments is picked-up from the bed and easily transported by the flow in the saltating mode. %
If turbulence is weak, the moving grains roll and slide along the bottom, interacting with the resting particles and the sediment transport rate is much smaller. %
%

%%
%%    Figure 14
%**********************************************************************
\begin{figure}
\begin{picture}(0,290)(0,0)
  \iffourteen
    \put(0,70){
    \put(-2,110){\includegraphics[trim=9cm 3cm 6cm 5cm, clip, width=.55\textwidth]{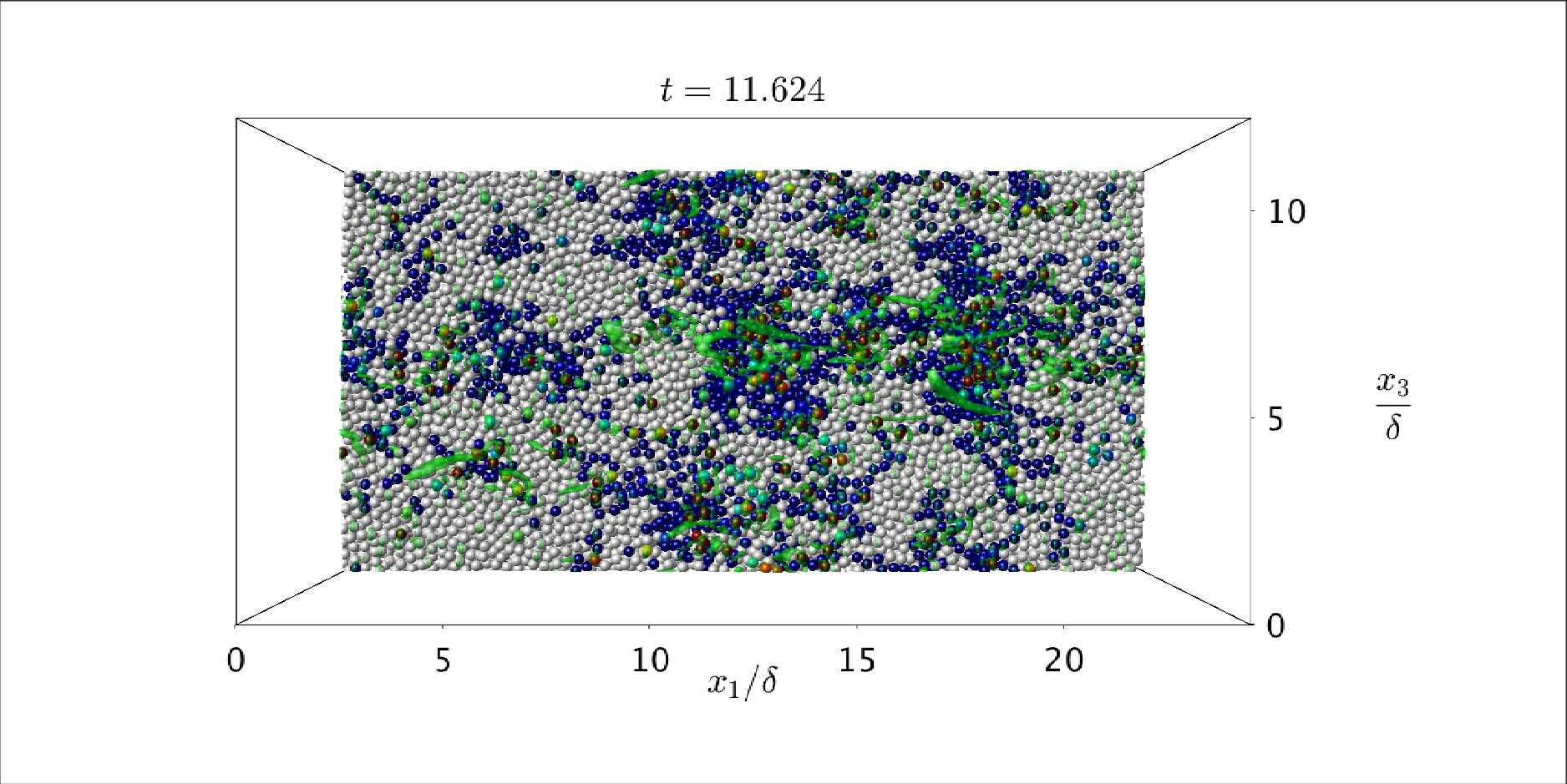}}
    \put(190,110){\includegraphics[trim=9cm 3cm 6cm 5cm, clip, width=.55\textwidth]{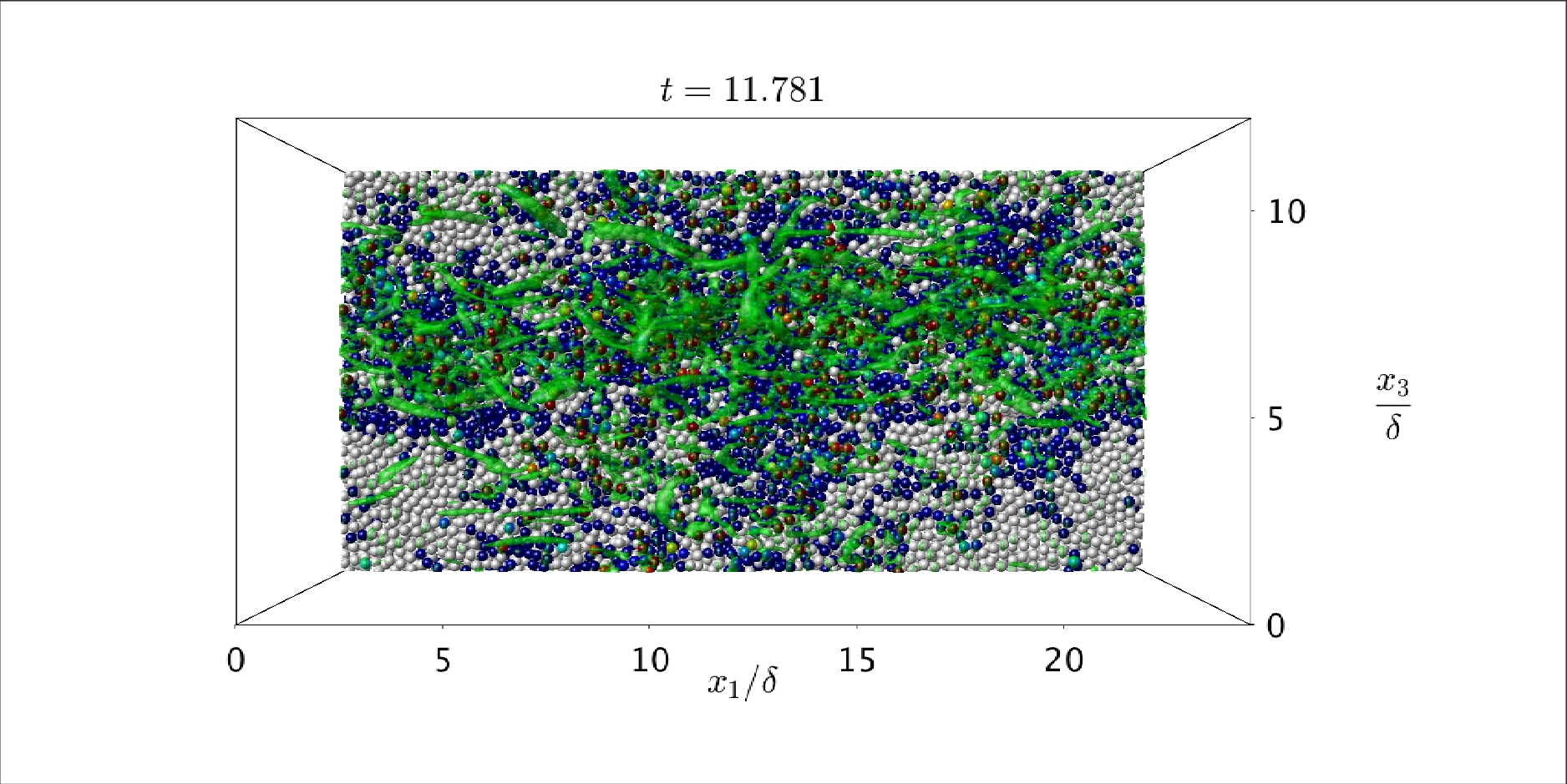}}
    \put(-2,0){\includegraphics[trim=9cm 3cm 6cm 5cm, clip, width=.55\textwidth]{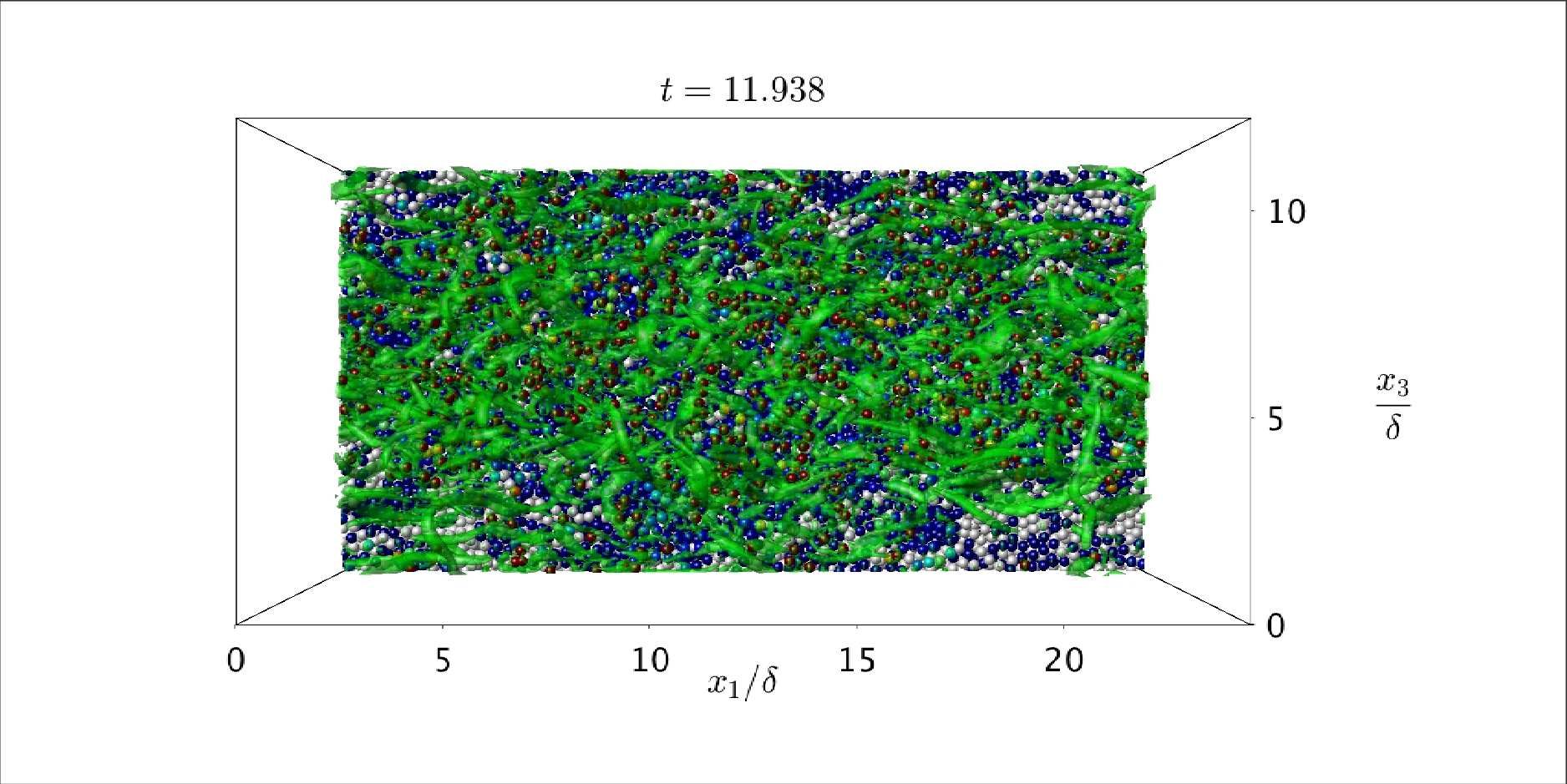}}
    \put(190,0){\includegraphics[trim=9cm 3cm 6cm 5cm, clip, width=.55\textwidth]{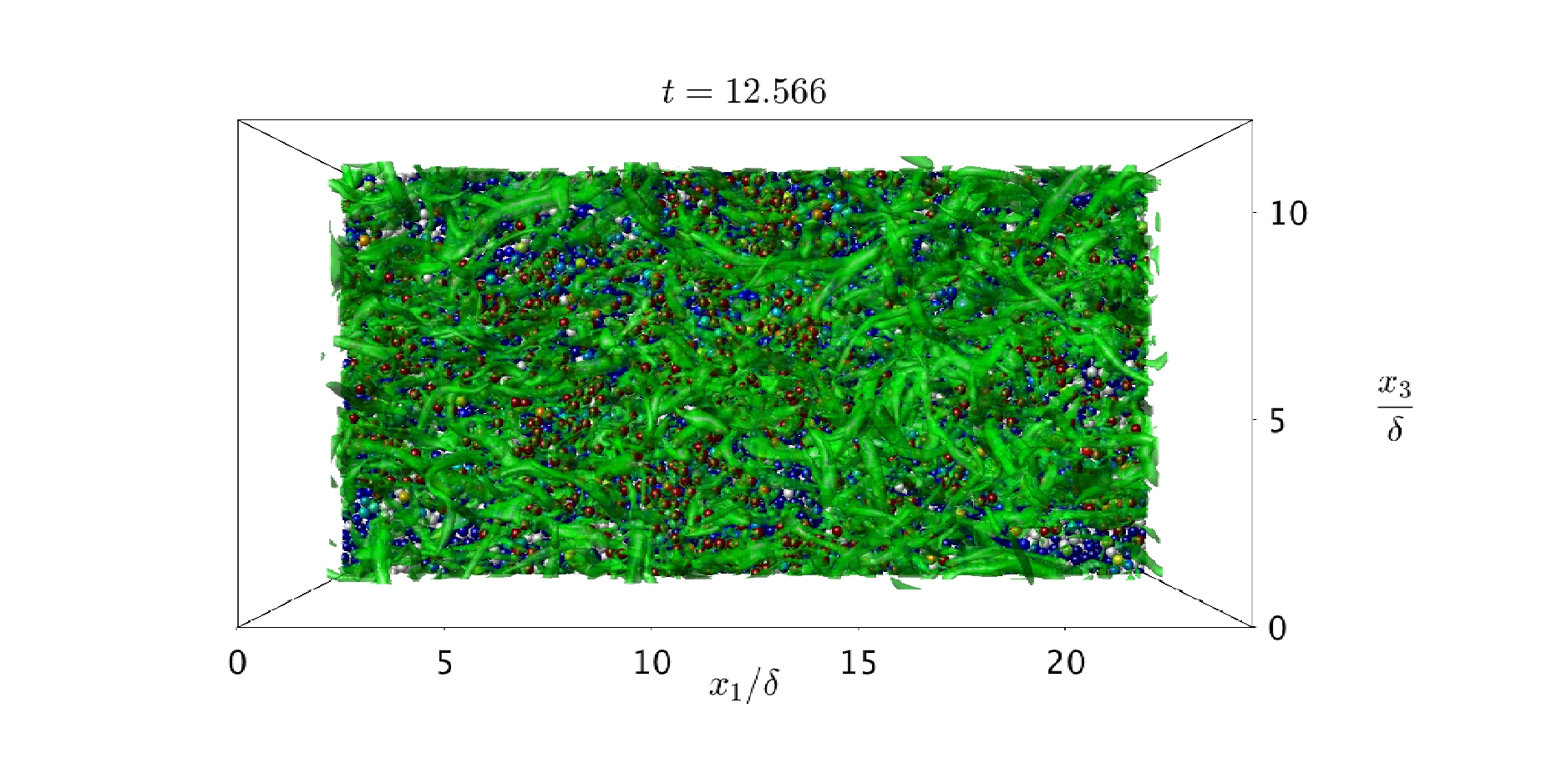}}
    \put(-2,0){
    \put(90,4){\tiny$^*$}
    \put(99,4){\tiny$^*$}
    \put(-8,98){$(c)$}
    }
    \put(190,0){
    \put(90,4){\tiny$^*$}
    \put(99,4){\tiny$^*$}
    \put(-8,98){$(d)$}
    }
    \put(-2,110){
    \put(90,4){\tiny$^*$}
    \put(99,4){\tiny$^*$}
    \put(-8,98){$(a)$}
    }
    \put(190,110){
    \put(90,4){\tiny$^*$}
    \put(99,4){\tiny$^*$}
    \put(-8,98){$(b)$}
    }
    \put(190,0){
    \put(208,56.5){\tiny$^*$}
    \put(208,48){\tiny$^*$}
    }
    \put(190,110){
    \put(208,56.5){\tiny$^*$}
    \put(208,48){\tiny$^*$}
    }
    } % put
    \put(0,-5){
    \put(90,0){\includegraphics[trim=0cm 0cm 0cm 0cm, clip, width=.5\textwidth]{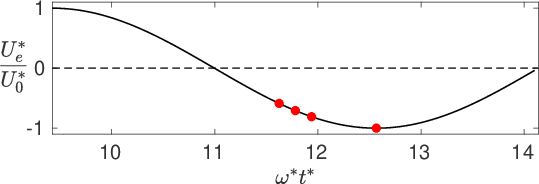}}
    \put(184,33){\scriptsize $(a)$}
    \put(189,19){\scriptsize $(b)$}
    \put(198,29){\scriptsize $(c)$}
    \put(219,25){\scriptsize $(d)$}
    }
  \fi
\end{picture}
\caption{%
Top view of the computational domain. Green surfaces visualise the isocontour of $\lambda_2^*\del^{*2}/\U^{*2}=-0.5$ at (from left to right and from top to bottom) $t = 3.70\pi$, $t = 3.75\pi$, $t = 3.80\pi$ and $t = 4.00\pi$. %
Spheres are coloured from blue to red on the basis of the magnitude of their instantaneous velocity ranging between $0$ and $0.1~\U^*$. %
White spheres are essentially at rest. %
$\Rdel=750$ and $d=0.335$ ($\runb$). %
A full sequence of similar top view visualisations can found online in the movie no.~$3$ of the \textit{supplementary material} which show isosurfaces of spanwise-vorticity fluctuations. %
}%
\label{fig14}
\end{figure}
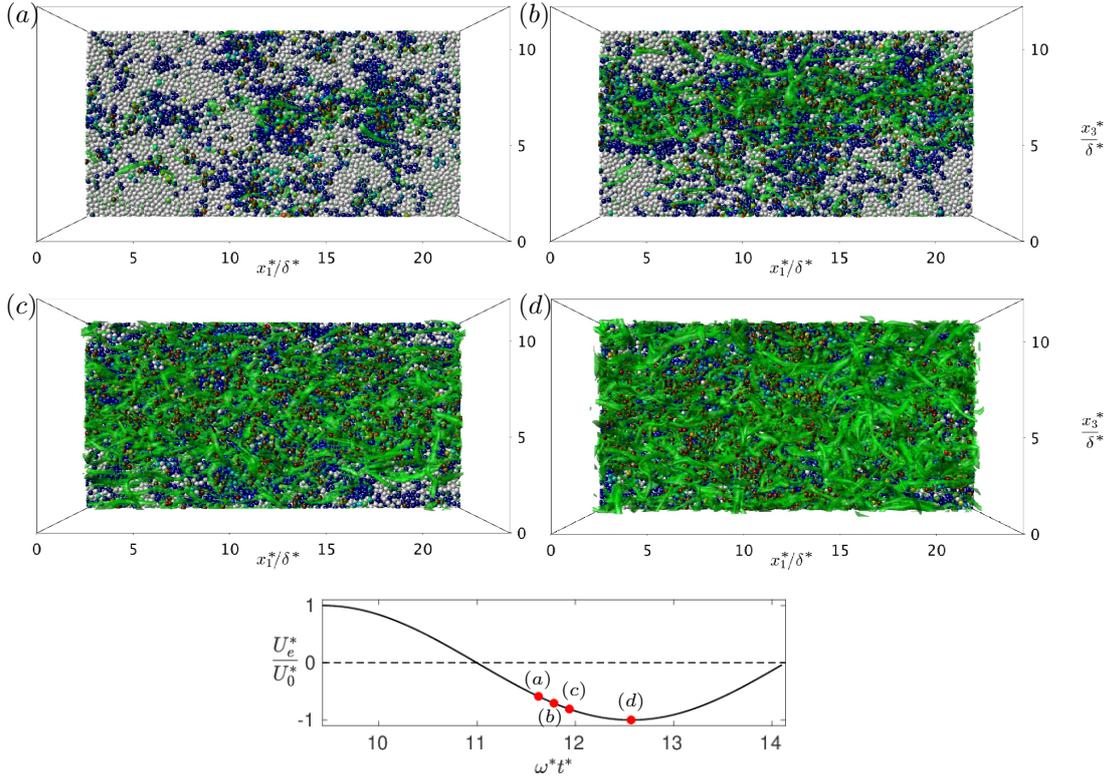
%**********************************************************************
%*****************************************************
%%
%%    Figure 15a
\begin{figure}
\begin{picture}(0,246)(0,0)
  \iffifteen
    \put(0,164){\includegraphics[trim=0cm 0cm 0cm 1cm, clip, width=1\textwidth]{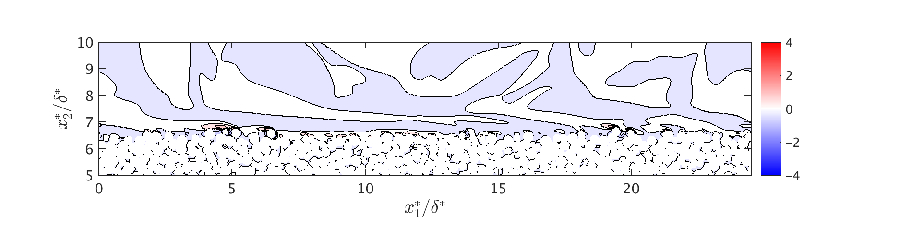}}
    \put(0,82){\includegraphics[trim=0cm 0cm 0cm 1cm, clip, width=1\textwidth]{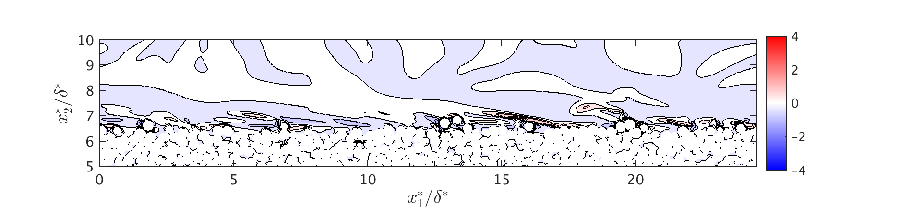}}
    \put(0,0){\includegraphics[trim=0cm 0cm 0cm 1cm, clip, width=1\textwidth]{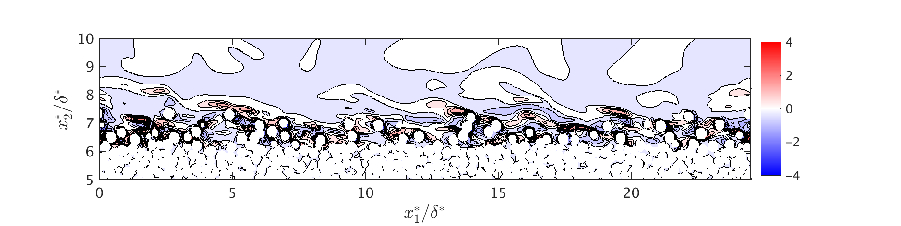}}
    \put(10,238){$(a)$}
    \put(10,156){$(b)$}
    \put(10,78){$(c)$}
  \fi
\end{picture}
\caption{%
Dimensionless fluctuating spanwise component of vorticity $\vort{3}^{'*}\del^*/\U^*=\vort{3}'=\vort{3}-\zxav{\vort{3}}$, in the plane $\xf{3}=6~\delta^*$ for $\Rdel=750$ and $d=0.335$ ($\runb$) at (a) $\omega^*t^*=3.6 \pi$, (b) $\omega^*t^*=3.7 \pi$, (c) $\omega^*t^*=3.8 \pi$. %
}%
\label{fig15a}
\end{figure}
%%
%%%%% Figure ??  *****************************************************
\begin{figure}
\begin{picture}(0,120)(0,0)
  \ifeighteen
  \put(-30,0){%
    \put(30,0){\includegraphics[trim=0cm 0cm 0cm 0cm, clip, width=.47\textwidth]{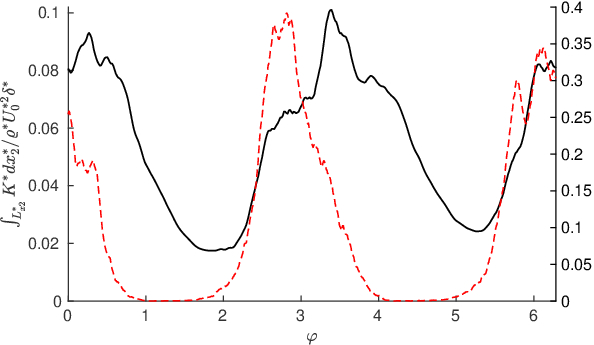}}
    \put(212,30){\rotatebox{90}{\tiny $\qpmean^*/\sqrt{(s-1)g^*d^{*3}}$}}
    \put(25,95){\small $(a)$}
   }%
   \put(170,0){%
    \put(30,0){\includegraphics[trim=0cm 0cm 0cm 0cm, clip, width=.47\textwidth]{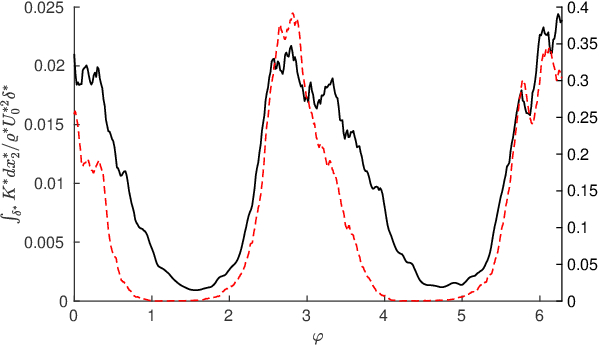}}
    \put(212,30){\rotatebox{90}{\tiny $\qpmean^*/\sqrt{(s-1)g^*d^{*3}}$}}
    \put(23,95){\small $(b)$}
   }%
  \fi
\end{picture}
\caption{%
Turbulent kinetic energy per unit bottom area (black continuous line), integrated along the $x_2$-direction $(a)$ from the bed surface to the top of the computational domain and $(b)$ within a layer $1~\del$ thick above the bed surface, plotted as a function of the phase $\varphi$ during the second cycle for $\Rdel=750$ and $d=0.335$ ($\runb$).%
The red broken lines indicate the dimensionless sediment flow rate. %
}%
\label{TEMPORARY5}
\end{figure}
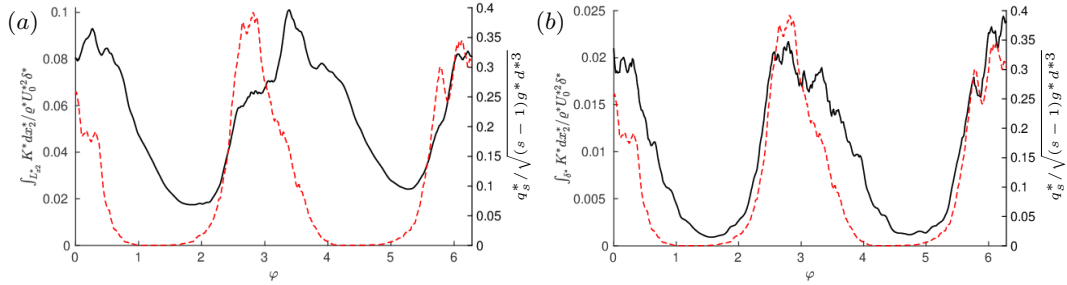
%%
%%%%% Figure 20  *****************************************************
\begin{figure}
\begin{picture}(0,190)(0,0)
  \iftwenty
    \put(30,0){\includegraphics[trim=0cm 0cm 0cm 0cm, clip, width=.8\textwidth]{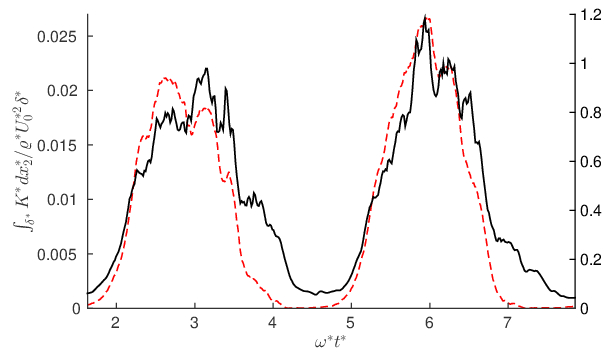}}
    \put(338,65){\rotatebox{90}{$\qpmean^*/\sqrt{(s-1)g^*d^{*3}}$}}
  \fi
\end{picture}
\caption{%
Turbulent kinetic energy per unit bottom area (black continuous line) integrated along the $x_2$-direction within a layer $1~\del$ thick above the bed surface, plotted as function of time for $\Rdel=1000$ and $d=0.335$ ($\rund$). %
The broken line indicates the dimensionless sediment flow rate. %
}%
\label{fig20}
\end{figure}
%
% FIGURE 20
%*****************************************************
\begin{figure}
\begin{picture}(0,340)(0,0)
  \iftwentythree
  \put(10,162){%
    \put(0,0){\includegraphics[trim=0cm 0cm 0cm 0cm, clip, width=.9\textwidth]{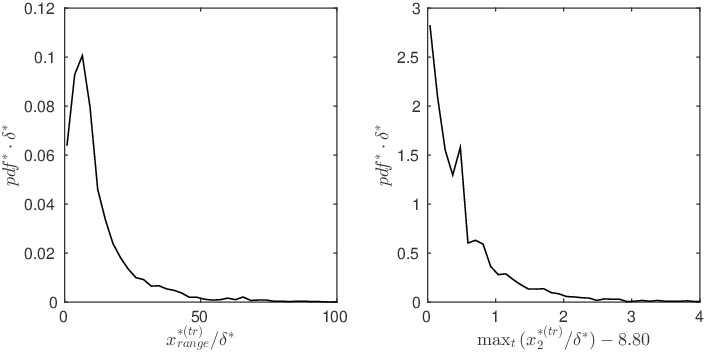}}
   }%
  \put(10,0){%
    \put(0,0){\includegraphics[trim=0cm 0cm 0cm 0cm, clip, width=.9\textwidth]{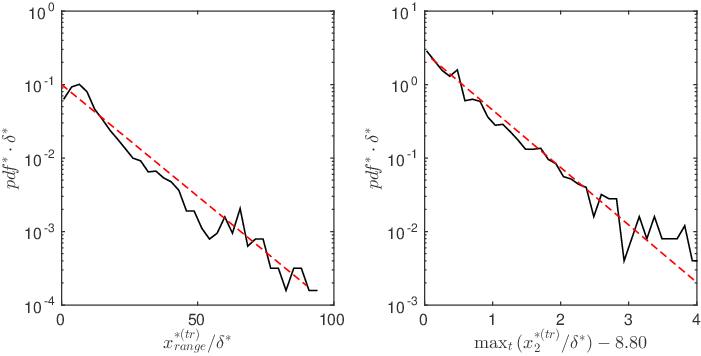}}
   }%
  \put(2,315){$(a)$}
  \put(190,315){$(b)$}
  \put(2,155){$(c)$}
  \put(190,155){$(d)$}
  \fi
\end{picture}
\caption{%
Panels $(a,c)$ show the pdf, in linear and semi-logarithmic scales, of the length of the jumps experienced by particles that cross the plane $x_2^*=\widehat{x}_2^*=8.80~\delta^*$. %
Similarly, $(b,d)$ show the pdf of the maximum height above the plane $x_2^*=\widehat{x}_2^*$ reached by particles. %
Broken red lines, that approximate the pdfs, correspond to the functions $(c)$: $pdf\cdot\delta=0.1\,\exp{(-0.07\,x_{range}^{*(tr)}/\delta^*)}$ and $(d)$: $pdf\cdot\delta=2.72\,\exp{(-1.8\,\max_t{(x_2^{*(tr)}/\delta^*)+15.84})}$. %
$\Rdel=1000$ and $d=0.335$ ($\rund$). %
}%
\label{fig23}
\end{figure}
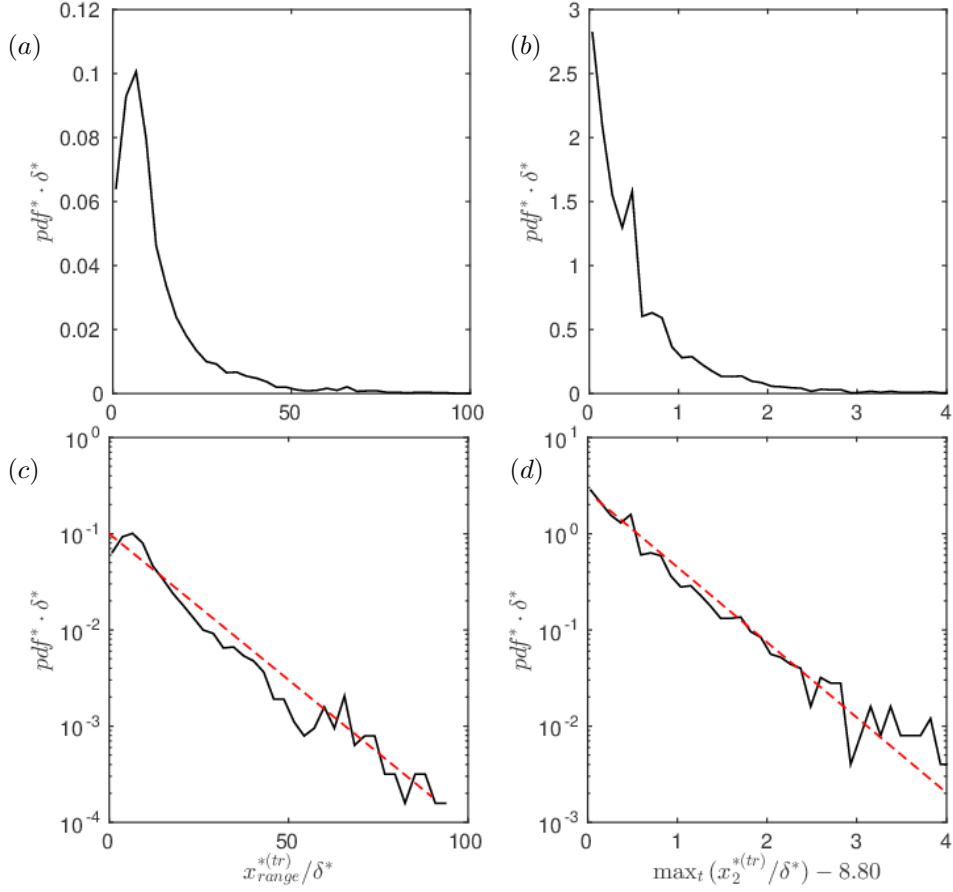
%*****************************************************

To relate the sediment motion to the dynamics of the coherent vortex structures and the turbulent eddies generated by the transition from the laminar to the turbulent regime, let us consider again the numerical simulation carried out for $\Rdel=750$ and $\ds=0.335$. %
Figure \ref{fig14} shows a top view of the bed where the sediment grains are coloured according to their velocity. %
Simultaneously, figure \ref{fig14} shows the coherent vortex structures which characterise the turbulent flow and are visualised by the $\lambda_2$-criterion \citep{JeongHussain1995}. %
The surfaces which appear in figure \ref{fig14} are characterised by a small negative value of the second eigenvalue $\lambda_2$ of the matrix $D^2+\Omega^2$, $D$ and $\Omega$ being the symmetric and antisymmetric parts of the gradient of the velocity field. %
The coherent vortex structures generated by turbulence appearance cause local high values of the bottom shear stress when they interact with the bottom and the sediment grains move with large velocities in the areas below the coherent vortices. %
In particular, it appears that patches of sediments randomly distributed over the bed are convected in the flow direction when coherent vortex structures are generated by turbulence appearance and move close to the bottom (see figure \ref{fig14}a). %
Then, at $\om^*t^*=3.75 \pi$, the turbulent eddies give rise to a band aligned with the $\xf{1}$-direction and almost in the center of the computational domain. Below the band of vortex structures, the sediment grains move with the largest velocities. %
Later, turbulence spreads over the whole domain and an intense sediment transport is observed over the entire bed. %
Eventually, turbulence decays, thus the vortex structures which characterised the flow field weaken and the sediment transport decreases till it becomes negligible at about $\om^*t^*=4.3 \pi$. %

These findings are further supported by figure \ref{fig15a} which shows that the sediment transport rate is largely affected not only by the external flow, but also by the interaction of the turbulent eddies with the sediment grains. %
At $\omega^*t^*=3.6 \pi$, turbulence is weak and only the grains in unstable positions slowly roll and move to attain more stable positions being dragged by the external flow. %
Then, turbulent fluctuations become intense, in particular
close to the bottom, and more sediments start to move because the fluctuating velocity components cause peaks of the hydrodynamic force acting on the sediment particles (see figure~\ref{fig15a}b). %
Later, the external velocity becomes larger as well as the turbulent eddies become more intense and the sediment grains not only move but they start to saltate, being picked-up from the bed. %

The time development of $K^*$ is shown in figure~\ref{TEMPORARY5}a where it appears that turbulence grows around $\varphi = 0.67 \pi$ and $1.67 \pi$ and then it attains its maximum value to slowly decrease later on and to assume relatively small values after flow inversion, thus being loosely related to the sediment transport which is also plotted in the same figure. %
Even though the growth of the turbulent kinetic energy and the growth of the sediment transport rate take place almost simultaneously, $K=\frac{K^*}{\rho^* U^{*2}_0\delta^*}$ attains its maximum value later than $q_s$ and it keeps large values even when $q_s$ vanishes. %
This finding can be easily understood taking into account that the sediment pick-up rate is mainly related to the upward velocity component generated by the turbulent eddies which are present close to the bottom. %
Figure~\ref{figY} shows that strong turbulent vortex structures are generated close to the bed when transition to turbulence takes place. %
However, later on, turbulence diffuses far from the bottom and it interacts no longer with the moving sediments which slow down and come to rest. Hence, the curve $q_s(t)$ follows more closely that obtained by considering the turbulent kinetic energy $K$ per unit area integrated from $\xf{2}^*=0$ up to the horizontal plane located at distance $\del^*$ from the instantaneous bottom surface (see figure~\ref{TEMPORARY5}b). %

If the Reynolds number $\Rdel$ is increased, turbulence strength increases as shown by figure~\ref{fig20} where $K$ is plotted versus the phase within the cycle for $R_\delta=1000$ ($\rund$). %
Because of the high turbulence intensity, a great number of particles is picked-up from the bed during the phases characterised by large values of the bottom shear stress. %
Once picked-up, some of the particles cover large distances without interacting with the bottom, before coming to rest again. %
This result is clearly shown by figure~\ref{fig23}a,c where the probability density function of the length of particle jumps is plotted for the particles that are moving above the plane $x_2^*-x_{2,bottom}^*=0.67~\delta^*=2.01~\ds^*$, suitably chosen to distinguish the saltating particles from the particles which roll and slide
on the resting particles. %
For $\rund$, during the oscillation period, the bottom elevation fluctuates of an amount approximately equal to $1~\ds^*$ above and $1~\ds^*$ below the time-average bottom elevation $x^*_{2,bottom}=8.13~\del^*$. %
Despite the fact that the average jump length is rather large, being about $36.2~\ds^*$, the median value is equal to $19.6~\ds^*$, since most of suspended particles rapidly re-deposit, as shown by figure~\ref{fig23}b,d, where the probability density function of the height of the particle jumps is plotted for the same particles considered in figure~\ref{fig23}a,c. %
Taking into account that the ratio $d^*/\delta^*$ is equal to $0.335$, the average height of the particle jumps turns out to be much smaller than the thickness of the region above the bottom where turbulence is intense (see for example figure~\ref{figY}) and it can be assumed that no significant suspended load is present. %
Indeed, the displacement boundary layer thickness for $\runb$, defined by equation~\eqref{eqdisp}, is $\del^*_{dis}=7.15~\del^*$. %
This numerical finding is consistent with the empirical criteria usually employed to determine the presence of sediment in suspension, e.g. \citet{bagnold1966,sumer2002}. %
The ratio $u_\tau^*/\vs^*$ between the shear velocity and the fall velocity of the particles ($\vs^*=\sqrt{(s-1)g^*d^*}$) is smaller than one and the Reynolds number, $u_{\tau,max}^*\ds^*/\nu^*$, is larger than $20$ when 
the maximum value of the bottom shear stress is considered (see table~\ref{tab0}). %
Indeed, the maximum value of $u_\tau^*/\vs^*$ for $\rund$ turns out to be about $0.6$. %

Notwithstanding the fact that the value of $K$ never vanishes and the random velocity fluctuations keep large during the whole flow cycle and tend to pick-up the sediments from the bottom, there are phases such that $q_s$ vanishes as shown by figure~\ref{fig20}. %
The vanishing of the sediment transport rate is related to the vanishing of the average bottom shear stress, which takes place twice during the cycle. %

\begin{figure}
\begin{picture}(0,180)(0,0)
  \iftwentyone
    \put(40,0){\includegraphics[trim=0 1cm 0 0, clip, width=0.75\textwidth]{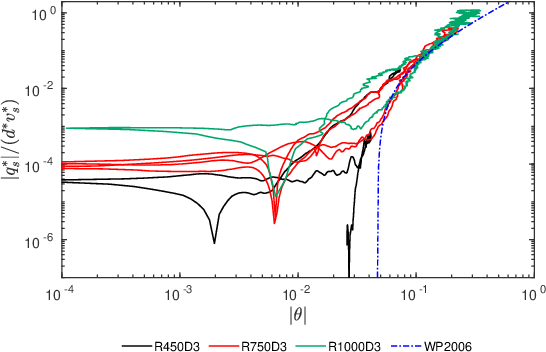}}
    \put(235,90){\rotatebox{60}{\vector(1,0){40}}}
    \put(220,140){\rotatebox{215}{\vector(1,0){30}}}
  \fi
\end{picture}
\caption{%
Dimensionless sediment flow rate $q_s$ as function of the absolute value of the Shields parameter for runs at $\Rdel=450$ (turbulent cycles), $750$ and $1000$ ($\runa$: black line, $\runb$: red line and $\rund$: green line, respectively). %
The blue dashed-dotted line represents the formula by \citet{wong2006}. %
Arrows indicate the orbital direction. %
}%
\label{fig21}
\end{figure}
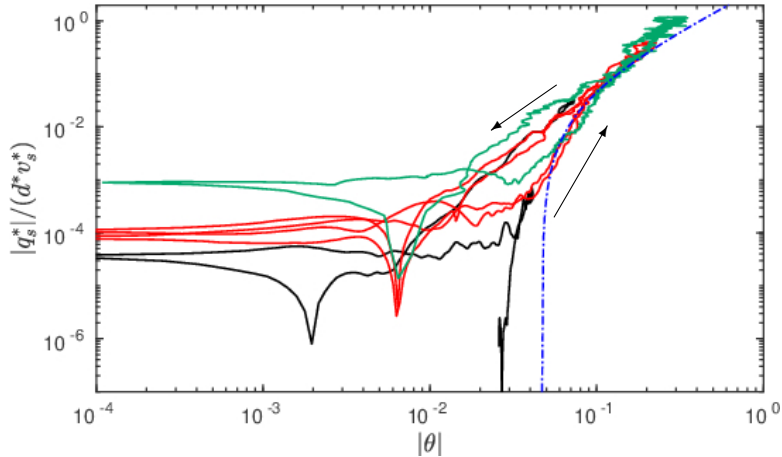

The results of the numerical simulations are summarised in figure~\ref{fig21} which shows the value of $q_s$ as a function of the Shields parameter $\shields$ for the three values of the Reynolds number presently considered. %
If small values of the Shields parameter are not considered, the present results suggest that fair predictions of the sediment transport rate can be obtained by looking for a correlation of $q_s$ with $\theta$ similar to that which describes $q_s$ versus $\theta$ in a steady flow. %
For example figure~\ref{fig21} shows that the relationship $q=a \left(\theta-\theta_{cr}\right)^{b}$, proposed by \citet{wong2006} to evaluate $q_s$ in steady flows with $a=4.93, \theta_{cr}=0.047, b=1.6$, provides results which fairly agree with those of the present simulations. %
When $\theta$ is close to $\theta_{cr}$ and $q_s$ is relatively small, it is necessary to take into account that the amount of sediment dragged by the fluid depends also on the time derivative of $\theta$. %
In fact for the same value of $\theta$, $q_s$ is larger if $d\theta/dt$ is negative. %
However, this hysteresis is present for such small values of $\theta$ that it can be neglected for practical applications. %
The fact that there is a rather small but finite sediment transport rate when $\shields$ tends to zero is due to the particle inertia and to the small effects of the imposed pressure gradient. %
For the same reason, the sediment transport rate tends to vanish for finite values of $\shields$ when particle inertia and the imposed pressure gradient effects balance the viscous force acting on sediment particles. %
%

%***********figure 21******************************************
%*****************************************************

%***********figure 22******************************************
\begin{figure}
\begin{picture}(0,180)(0,0)
  \iftwentytwo
    \put(40,0){\includegraphics[trim=0 1cm 0 0, clip, width=0.77\textwidth]{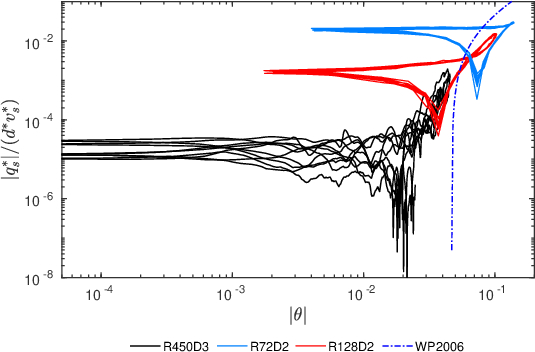}}
    \put(25,0){
    \put(265,142){\rotatebox{-112}{\vector(1,0){30}}}
    \put(284,160){\rotatebox{-112}{\vector(1,0){30}}}
    \put(200,146){\rotatebox{5}{\vector(1,0){50}}}
    \put(218,168){\rotatebox{2}{\vector(1,0){50}}}
    }
  \fi
\end{picture}
\caption{%
Dimensionless sediment flow rate $q_s$ as function of the absolute value of the Shields parameter for simulations at $\Rdel=72,\,128$ and $\ds\simeq 0.25$ (solid blue and red lines) \citep[non turbulent,][]{mazzuoli2019} and for the non-turbulent cycles of \runa at $\Rdel=450$ and $\ds=0.335$ (black line). %
The blue dashed-dotted blue line represents the formula by \citet{wong2006}. %
Arrows indicate the orbital direction. %
}%
\label{fig22}
\end{figure}
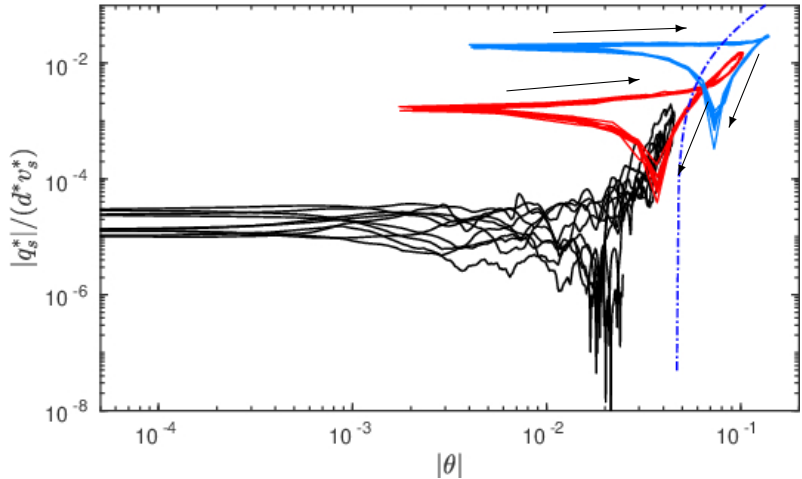
%*****************************************************

\section{Conclusions}
\label{conc}

New and interesting information of the sediment transport generated by sea waves is obtained by means of the DNSs which allow to evaluate the hydrodynamics within the oscillatory boundary layer generated by surface waves close to the bottom and to determine the dynamics of idealised sediment particles dragged by the flowing fluid. %

The values of the flow Reynolds number fall in the intermittently turbulent regime, such that turbulence is significant only during part of the flow cycle. %
The other parameters are typical of medium sand. %
Hence, the results are useful to quantify the bedload sediment transport outside the breaking and surf regions where higher values of the Reynolds number are usually found such that the DNSs of the turbulent flow field within the bottom boundary layer are presently unaffordable. %

The main result of the investigation is the description of sediment dynamics under the action of the turbulent eddies which are generated within the boundary layer. %
The pressure fluctuations induced by the turbulent eddies penetrate within the porous bed and generate lift forces which superimpose to those due to the pressure difference between the bottom and the top of the sediment particles that, in turn, is associated with the shear flow close to the bed surface. %
On average, the lift force due to the turbulent pressure fluctuations is upward directed and the sediment grains tend to be picked-up from the bed and then transported by the external flow in the saltation mode. %
On the other hand, when the flow re-laminarises but the bed shear stress is large enough to induce sediment transport, the sediment grains tend to roll and slide one over the top of the others. %
This particle dynamics is typical of a laminar flow and it gives rise to sediment transport rates quite different from those observed when turbulence is present. %

The differences between the values of $\qpmean$ generated by a laminar and a turbulent oscillatory boundary layer can be easily appreciated if the results of figure~\ref{fig21} are compared with those obtained by \citet{mazzuoli2019} which, for reader's convenience, are plotted in figure~\ref{fig22}. %
\citet{mazzuoli2019} investigated the formation of sea ripples by means of DNSs and computed the sediment transport rate for values of the parameters as those of some of the laboratory experiments of Blondeaux et al. (1988). %
In particular, the experiments characterised by $\Rdel=72$ and $\Rdel=128$ and by $\ds\simeq 0.25$ were considered by \citet{mazzuoli2019}. %
For such values of the Reynolds number, the flow regime is laminar. %
In figure~\ref{fig22}, the results obtained for $\Rdel=450$ and $\ds=0.335$ ($\runa$) during the half-cycles characterised by weak turbulence are also plotted. %
As already pointed out, in these cases, the values of $\qpmean$ during the accelerating phases are different from those computed during the decelerating phases, even if the Shields parameter $\shields$ is the same, because particle dynamics is affected not only by the bottom shear stress but also by the streamwise pressure gradient. %

In the turbulent regime, the bedload sediment transport rate observed for large values of $\shields$ during the accelerating phases is practically equal to that observed during the decelerating phases because the pressure gradient plays a negligible role in particle dynamics. %
In fact, the magnitude of the sediment transport rate during the accelerating phases differs from that during the decelerating phases only when the Shields parameter is quite small, the differences being mainly due to the different values of the turbulence intensity observed during the accelerating and decelerating phases, even for the same value of the bottom shear stress. %
The reader should notice that, when the flow regime is laminar, the sediment flow rate decreases if the Reynolds number is increased while, if turbulence is present, the values of $\qpmean$ increase if the Reynolds number is increased. %

Finally it is worth pointing out that, in the turbulent regime, fair predictions of the bedload sediment flux can be obtained by means of empirical formulae obtained on the basis of experimental measurements carried out in steady flow, at least for the typical periods of sea waves. %
Indeed, for high values of the Reynolds number, the amplitude of the fluid displacement oscillations turns out to be much larger than the grain size. %
Hence, the Keulegan-Carpenter number $\Kc$ of the flow around sediment grains is large and the sediment particles feel a succession of quasi-steady flows. %

\bigskip
This study has been funded by the Office of Naval Research (U.S.A.) (under the research project n. N62909-17-1-2144). %
The support of CINECA, who provided computational resources on Marconi under the PRACE project MOST SEA (Proposal ID: 2017174199), and a grant of computer time from the DoD High Performance Computing Modernization Program at the ERDC DSRC are also acknowledged. \\%
We also acknowledge that this research was also supported by the MIUR, in the framework of PRIN 2017, with grant number 20172B7MY9, project ``FUNdamentals of BREAKing wave-induced boundary dynamics''. %

Declaration of Interests. The authors report no conflict of interest.
\bibliographystyle{elsarticle-harv} 
\bibliography{\refdir ref}

\end{document}